\documentclass[12pt,onecolumn,draftclsnofoot]{IEEEtran}

\usepackage{setspace}
\usepackage{clrscode}
\usepackage{amsmath}
\usepackage{amssymb}
\usepackage[dvips]{graphicx}
\usepackage{epsfig}
\usepackage{hyperref}
\usepackage{bm}
\usepackage{subfigure}
\usepackage{algorithm}
\usepackage{algorithmic}
\usepackage{setspace}
\begin{document}

\title{How to Minimize the Weighted Sum AoI in Multi-Source Status Update Systems: OMA or NOMA?}
    \author{
\IEEEauthorblockN{Jixuan Wang and Deli Qiao}
\thanks{This work is supported in part by the Shanghai Rising-Star Program (21QA1402700).}
\thanks{The authors are with the School of Communication and Electronic Engineering, East China Normal University, Shanghai, China (email:51205904028@stu.ecnu.edu.cn, dlqiao@ce.ecnu.edu.cn).}
}

\maketitle

\begin{abstract}
    In this paper, the minimization of the weighted sum average age of information (AoI) in a multi-source status update communication system is studied. Multiple independent sources send update packets to a common destination node in a time-slotted manner under the limit of  maximum retransmission rounds. Different multiple access schemes, i.e., orthogonal multiple access (OMA) and non-orthogonal multiple access (NOMA) are exploited here over a block-fading multiple access channel (MAC). Constrained Markov decision process (CMDP) problems are formulated to describe the AoI minimization problems considering both transmission schemes. The Lagrangian method is utilised to convert CMDP problems to unconstraint Markov decision process (MDP) problems and corresponding algorithms to derive the power allocation policies are obtained. On the other hand, for the case of unknown environments, two online reinforcement learning approaches considering both multiple access schemes are proposed to achieve near-optimal age performance. Numerical simulations validate the improvement of the proposed policy in terms of weighted sum AoI compared to the fixed power transmission policy, and illustrate that NOMA is more favorable in case of larger packet size.
\end{abstract}

\section{Introduction}

In many emerging real-time Internet of Things (IoT) applications \cite{L. D. Xu}, \cite{T. Qiu}, particularly for those dealing with time-sensitive data, e.g., autonomous driving, intelligent traffic-monitoring networks and disaster monitoring and alerting systems, strictly guaranteeing the timeliness of information updates is crucial since outdated information might become worthless. From the perspective of system, the knowledge of the status of a remote sensor or system requires to be as timely as possible, so the timeliness of state updates has evolved into a new field of network research \cite{R. D. Yates}. To characterize such information timeliness and freshness, the metric termed age of information (AoI), typically defined as the time elapsed since the most recent successfully received system information was generated at the source, has been proposed \cite{S. Kaul}.

Most of the earlier work on AoI in various networks mainly consider simple single-source single-destination status update system models (see, e.g., \cite{S. Kaul}-\cite{Y. Sun}), while recent researches related to AoI optimization have shifted to more practical multi-source and/or multi-destination systems and most of them involve orthogonal multiple access (OMA) technique \cite{H. Tang}-\cite{I. Kadota}. For instance, the authors in \cite{H. Tang} considered a system model in which a central controller collects data from multiple sensors via wireless links and the AoI optimization problem is subject to both bandwidth and power consumption constraints. Besides, in \cite{H. Tang}, a truncated scheduling policy was proposed to satisfy the hard bandwidth constraint. The work in \cite{W. Pan} presented two multi-source information update problems in a practical IoT system, called AoI-aware Multi-Source Information Updating (AoI-MSIU) and AoI-Reduction-aware Multi-Source Information Updating (AoIR-MSIU) problems,respectively. A wireless broadcast network with random arrivals was considered in \cite{Y. -P. Hsu}, where two offline and two online scheduling algorithms were proposed, leveraging Markov decision process (MDP) techniques and the Whittle's methodology for restless bandits. Similarly, the system in \cite{Z. Jiang} was also with stochastic arrivals, and the Whittle's index policy with near-optimal performance was derived in closed-form. In multi-user multi-channel systems \cite{Z. Qian}, the authors proposed two asymptotic regimes to investigate how to exploit multi-channel flexibility to improve the age performance. In \cite{I. Kadota}, to minimize the expected weighted sum AoI subject to minimum throughput requirements, the authors developed four low-complexity scheduling policies and then compared their performance to the optimal policy.

Even though the literature mentioned above is all about OMA technique, non-orthogonal multiple access (NOMA) is a promising technique to reduce the average AoI \cite{A. Maatouk}-\cite{J. Chen} by using Successive Interference Cancellation (SIC), which will greatly improve spectral efficiency compared to OMA in large-scale system. In \cite{A. Maatouk}, the authors analyzed the performance of NOMA in minimizing AoI of a two-node uplink network for the first time and the results showed that OMA and NOMA can outperform each other in different configurations. A hybrid NOMA/OMA scheme was proposed in \cite{Q. Wang2}, in which the BS can adaptively switch between NOMA and OMA for the downlink transmission to minimize the AoI and a suboptimal policy called action elimination with lower computation complexity was also proposed. A uplink NOMA system combining  physical-layer network coding (PNC) and multiuser decoding (MUD) was considered in \cite{H. Pan}. In \cite{J. Chen}, the authors proposed an an adaptive AoI-aware buffer-aided transmission scheme (ABTS) on downlink communication system to improve AoI performance of a downlink NOMA system. However, these articles do not consider retransmission mechanism, which is essential to guarantee a certain degree of reliability of the received update packets.

Furthermore, dealing with age-optimal scheduling problem using reinforcement learning approaches in an unknown environment has recently drawn great attention \cite{E. T. Ceran1},\cite{E. T. Ceran2}-\cite{M. Akbari} and to the best of our knowledge, the first application of RL approaches to the problem with a minimum AoI criterion appeared in \cite{E. T. Ceran1}, which employed the \emph{average-cost} SARSA with \emph{softmax} algorithm to learn the system parameters and the transmission policy under hybrid ARQ (HARQ) protocols. As an extension of \cite{E. T. Ceran1}, in \cite{E. T. Ceran2}, the age-optimal problem was extended to a multi-user setting in the downlink system with orthogonal transmissions and three different RL methods were proposed to provide near-optimal performance. The work in \cite{H. Huang} proposed one off-line power control policy and two online RL algorithms to investigate the joint optimization considering both AoI and total energy consumption in the fading channel and now we extend this work to multi-source scenarios. The Policy Gradients and Deep Q-learning(DQN) methods were introduced in \cite{H. B. Beytur} to address a multi-queue AoI-optimal scheduling problem. An AoI-based trajectory planning (A-TP) algorithm employing deep reinforcement learning (DRL) technique was proposed in \cite{C. Zhou} to solve the online AoI-based trajectory planning problem in UAV-assisted IoT networks. In \cite{A. A. Al-Habob}, an underwater linear network was considered, in which the authors developed an actor-critic DRL based on a deep deterministic policy gradient (DDPG) method to minimize the normalized weighted sum AoI. In \cite{M. Akbari}, the authors developed single-agent and cooperative multi-agent virtual network function (VNF) placement utilizing DRL method to minimize VNF placement cost, scheduling cost, and average AoI in industrial internet of things (IIoT). However, none of the multi-user system works consider NOMA transmission scheme when using RL to solve AoI minimization problems in the unknown environment. Motivated by this, this paper makes the first attempt, to the best of our knowledge, to propose two RL algorithms considering both OMA and NOMA schemes in a multi-source wireless uplink status update system with unknown environment, and achieves the near-optimal age performance compared to the known environment.

In this paper, we investigate the weighted sum average AoI minimization problems for both OMA and NOMA schemes, where we consider not only the retransmission mechanism on the uplink communication system with a limit on the maximum number of retransmission rounds, but also the fact that each source is subject to individual power constraint. On the one hand, when the channel distribution information (CDI) is available at the transmitters, we formulate the AoI minimization problems as CMDP problems. Through the Lagrangian method, we relax the CMDP problems into equivalent MDP problems and accordingly obtain the algorithms to derive optimal policy in OMA and NOMA, respectively. Then on the other hand, we consider transmitting update packets over an unknown environment in which case we propose two RL algorithms to find the optimal power control policy. The main contributions of this article can be summarized as follows:
\begin{itemize}
  \item  Unlike most previous works where transmit power is fixed, we assume different transmit power levels in this paper, varying with the instantaneous AoI and transmission rounds. As a result, power control policy is derived to address the AoI-optimal problems.
  \item  When the CDI is available at the transmitters in advance, we first propose two off-line Value Iteration Algorithms (VIAs) to solve the Bellman optimality equations for both OMA and NOMA. Then with the minimization of \emph{value function}, the optimal policy is derived to achieve the AoI-optimal performance under the average power constraints. 
  \item  When such environment is not known as a priori, we design two online Q-learning with $\epsilon$-greedy exploration algorithms to find the optimal policy on both OMA and NOMA schemes.
  \item Numerical results show the comparison of AoI performance of OMA and NOMA and verify their performance improvements over fixed power schemes under the proposed policy. And RL-based approaches are shown to minimize weighted sum AoI in OMA and NOMA without degrading the performance significantly.
\end{itemize}

The remainder of this paper is organized as follows. In Section II, we discuss the preliminaries related to system model and AoI. Section III formulates the CMDP optimization problems. Section IV presents the details of the solutions to the optimization problems, and section V describes reinforcement learning approaches. Numerical results are provided in Section VI. Finally, section VII concludes the paper.

\section{Preliminaries}

\subsection{System Model}
\begin{figure}[h]
    \centering
    \includegraphics[width=0.5\textwidth]{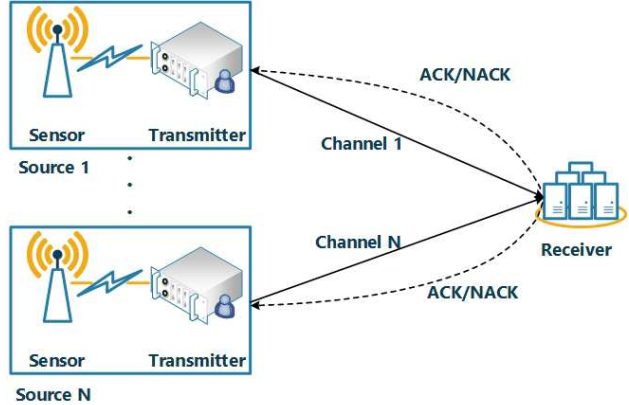}
    \caption{System model.}
    \label{fig:system model}
\end{figure}
In this paper, we consider a multi-source wireless uplink system as depicted in Fig. \ref{fig:system model}, where $N$ sources send their status update packets to a common receiver. Time is divided into slots of period $T$, and a so-called generate-at-will model is adopted where each sensor can generate a new status update at the beginning of any time slot. Additionally, $N$ independent error-free and delay-free feed back channels from destination to each source are considered. Here we assume that an update packet of $R$ bits is assigned to be transmitted in each slot and the receiver will send a ACK as feedback to the source if the sent package is successfully decoded such that the source is able to generate a new package that contains the latest status information with a time stamp. In case of decoding failure, a NACK from destination to source will be sent to request another round of transmission of the same packet. It is worth noting that one update packet can be transmitted up to $M$ times to ensure a certain degree of reliability. Namely, when reaching the maximum number of transmissions $M$, this packet should be discarded and a new packet will be generated and sent to the destination. In this model, we consider a block-fading MAC where the channel gain remains constant in each slot and varies independently over different time slots. For simplicity, $T$ is assumed to be 1 in the paper. Some details about OMA and NOMA are as follows:
\begin{enumerate}
  \item \textbf{Orthogonal multiple access (OMA)}: We assume that each source is assigned a single orthogonal block that is free of any interference from the others. Assume that the $n$-th $(n\in[N] \triangleq \{1,2,\ldots,N\})$ source occupies $\rho_n$ fraction of one time slot to transmit update packet. We should have $\sum_{n=1}^{N} \rho_n \leq 1$ and $\rho_n \geq 0, \forall n \in [N].$ As a consequence, by dividing the time slot, every source has opportunity to transmit their update packets without interfering with each other in a time slot.
  \item \textbf{Non-orthogonal multiple access (NOMA)}: On the other hand, NOMA allows every source to transmit update packets simultaneously to a common receiver with non-orthogonal signaling. It is clear that some signal interference exists between the wireless communication links. So when NOMA is conducted in time slot $t$, through SIC, the sources are allocated different power levels according to assigned decoding order for distinguishing between the sources and then successfully recover their corresponding information at the receiver in one time slot.
\end{enumerate}

\subsection{Age of Information(AoI)}
\begin{figure}
    \centering
    \includegraphics[width=0.4\textwidth]{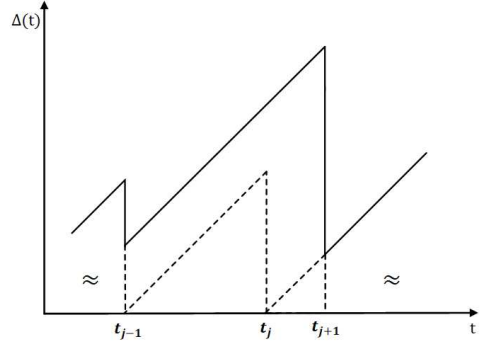}
    \caption{Evolution of age of information.}
    \label{fig:Evolution of age of information}
\end{figure}

Age of information (AoI) characterizes the timeliness and freshness of the information on the status update system. It is defined as the time elapsed since the most recent successfully received update was generated at the source. Suppose that at any time $t$, the last successfully received packet has a time stamp $U(t)$ representing its generation time. Then the age can be defined as
\begin{small}
\begin{align}
\Delta(t)= t -U(t).
\end{align}
\end{small}
The evolution process of the AoI $\Delta(t)$ is illustrated in Fig. \ref{fig:Evolution of age of information}, where $t_{j}$ denotes the time instance when the $j$-th packet is generated. In the figure, the solid lines denote the evolution of the age, while the dashed lines depict the age of each packet. When the currently transmitted package is successfully received, the age is reduced and a new packet will be generated. On the other hand, if a packet fails when the time interval between two consecutive packet generations $t_{j} - t_{j-1}$ equals $MT$, i.e., the number of transmission rounds reaches $M$, the age will keep increasing linearly until another successful reception. Just like in this figure, the $(j-1)$-th packet fails at the $M$-th transmission round, while the $j$-th packet succeeds. The average age of information for a single source is given by
\begin{small}
\begin{align}
\overline{\Delta}_{n} = \lim_{\tau \to \infty} \frac{1}{\tau} \int_{0}^{\tau}\Delta_{n}(t)\mathrm{d}t, \quad n = 1,2,\ldots,N\label{average AoI}
\end{align}
\end{small}where $\overline{\Delta}_{n}$ represent the average AoI of source $n$. Since AoI can be measured in terms of time slots, the integration in (\ref{average AoI}) can be described as the area under the sawtooth curve in Fig.\ref{fig:Evolution of age of information}, which is also the sum of multiple trapezoidal areas. So now the expression (\ref{average AoI}) can be rewritten as follows:
\begin{small}
\begin{align}\label{improved average AoI}
\overline{\Delta}_{n} = \lim_{J \to \infty} \frac{1}{T} \cdot \frac{\sum_{j=1}^{J}(2\Delta_n(t)+T)T}{2J}, \quad n = 1,2,\ldots,N
\end{align}
\end{small}where $J$ is the total number of slotted trapezoids. Recalling that the duration of one time slot is normalized as $T = 1$, so, specially, the expression (\ref{improved average AoI}) can be simplified as $\frac{\Delta_{n}(t)+\Delta_{n}(t)+1}{2} = \Delta_{n}(t) + \frac{1}{2}$.
\section{CMDP Problem Formulation}

In this section, we first clarify five tuples in the CMDP problems: states, actions, transition probabilities, rewards and costs. Subsequently we formulate CMDP problems that minimize the weighted sum average AoI subject to the individual power constraint at each source.
\subsection{System States}

In this paper, we define $s(t) = \left(m_{1}(t),\Delta_{1}(t);m_{2}(t),\Delta_{2}(t);\ldots;m_{N}(t),\Delta_{N}(t)\right)$ as the system state at any time $t$, which consists of the transmission round and the instantaneous AOI of each source, respectively. The state space $\mathcal{S} \subset \left(\{1,\cdots,M\} \times \mathbb{Z}^{+}\right )^{N}$ is in general countably infinite. It is worth noting that the transmission round $m$ is always logically less than or equal to the instantaneous AoI $\Delta$, and thus those points with $m > \Delta$ are supposed to excluded from the state space to ensure that the system is always feasible. 

Assume that the channel distribution information is available only at each source. Due to the block-fading assumption, each channel is modeled as a quantized channel where the channel gain set is denoted as $\digamma = \{z_{0}, z_{1}, \ldots, z_{K}\}$ in increasing order with $z_{0} = 0$ and $z_{K} = \infty$. ${K}$ is the quantization level. The channel is assumed to be in state $i$ when channel gain $z \in [z_{i},z_{i+1})$ and we assume that each channel is independently and identically distributed (i.i.d.) such that the probability of channel state transition from state $j$ to state $i$ is not relevant to the state transitions and is given by
\begin{small}
\begin{align}
\label{StateProbability}
{\rm Pr}\{z_{i}\mid z_{j}\} = {\rm Pr}\{z_{i}\} =\int_{z_{i}}^{z_{i+1}}p_{z}(z)\mathrm {d}z = \psi_{i},
\end{align}
\end{small}where $p_{z}(z)$ is probability density function of the channel gain $z$ and $\psi_{i}$ is the probability of channel state $i$. 
\subsection{Actions}
Note that both orthogonal transmission and non-orthogonal transmission are considered in the paper. So we will discuss actions in two parts.
\subsubsection{Actions in OMA}
In this scenario, one time slot is divided into $N$ parts to be assigned to $N$ sources. For each system state, the sources can choose different transmit power values (i.e., take different actions $a(m_n,\Delta_n)= P_n$) to send update information.

We assume that the additive Gaussian noise is of unit variance and the bandwidth is one. Therefore, for example, if $\rho_n$ fraction of one time slot is allocated to source $n$, the instantaneous channel rate of source $n$ with channel gain $z_{i}$ is given by $\rho_n {\rm log_{2}}(1+\frac{P_{n,i}}{\rho_n}z_{i})$ with boosted transmit power $\frac{P_{n,i}}{\rho_n}$, so the $i$-th transmit power level of source $n$ in OMA can be specified by following formula:
\begin{small}
\begin{align}
\label{OMA Action1}
P_{n,i}^{{\rm O}} = \rho_n \frac{{\rm 2}^{\frac{R^{\rm{O}}_n}{\rho_n}}-1}{z_{i}},\quad i = 1,2,\ldots,K,\,\forall n \in [N]
\end{align}
\end{small}where $R^{\rm{O}}_n$ represents the date rate or packet size equivalently of source 1 following OMA transmission scheme and the values of $\{z_{i}\}$ can be calculated from (\ref{StateProbability}) with the given values of $\pmb{\psi} = (\psi_{0},\psi_{1},\ldots,\psi_{K-1})$ accordingly.
For arbitrary source, when the action $a(m_{n},\Delta_{n}) = P_{n,i}^{{\rm O}}$ is taken, according to the independent block-fading assumption, the probability of erroneous transmission in OMA can be calculated as follows:
\begin{small}
\begin{align}
\label{OMA errorpr}
\epsilon_{n,i}^{{\rm O}} = {\rm Pr}\{R^{\rm{O}}_n<\overline{R}^{\rm O}\}={\rm Pr}\{z<z_{i}\} = \sum_{m=0}^{i-1}{\rm Pr}\{z_{m}\leq z <z_{m+1}\} = \sum_{m=0}^{i-1}\psi_{m},
\end{align}
\end{small}where $\overline{R}^{\rm O}$ is the minimum preset rate required to meet the transmission requirements which is set to be same for all sources.
\subsubsection{Actions in NOMA}
In the case of NOMA, the signals from different sources will interfere with each other, resulting in a high probability of erroneous transmission. With the aim of decoding the corresponding information correctly at the receiver so as to minimize the weighted sum average AoI, the SIC technique is adopted here. When $N$ sources transmit update packets, they can be decoded in $N!$ different orders. Let $\pmb{\mathcal{D}}=\{\mathcal{D}_1(\cdot),\mathcal{D}_2(\cdot),\ldots,\mathcal{D}_{N!}(\cdot)\}$ denote the decoding order set, where $\mathcal{D}_i(k)$ represents the original index of the source that ranks $k$-th in decoding order $\mathcal{D}_i$. Consequently, for each system state, the sources can choose different transmit power values, and the receiver can choose different decoding orders.

Taking decoding order $\mathcal{D}_i$ as an example, the message from source $\mathcal{D}_i(1)$ is scheduled to be decoded and recovered first when the decoding order is $\mathcal{D}_i$ and is assigned a strongest transmit power, while other sources are allocated weaker transmit power. Thus the information from other sources is buried in received signal and would be treated as interference noise by the common receiver. Based on the channel statistics, the transmit power set can be given by $\pmb{P^{{\rm N},\mathcal{D}_i}}$, consisting of $N*K$ elements, where the $(n*k)$-th element of $\pmb{P^{{\rm N},\mathcal{D}_i}}$, i.e., $P_{n,k}^{{\rm N},\mathcal{D}_i}$, represents the transmit power of source $n$ in state $k$ when decoding order is $\mathcal{D}_i$ considering NOMA case. With the actions set $\pmb{P^{{\rm N},\mathcal{D}_i}}$, the information rate for source $\mathcal{D}_i(n)$, denoted by $R^{{\rm{N}},\mathcal{D}_i}_{\mathcal{D}_i(n)}$, is given by
\begin{small}
\begin{align}
\label{NOMA Action1 X}
R_{\mathcal{D}_i(n)}^{{\rm{N}},\mathcal{D}_i} = \log_2\left(1+\frac{P_{\mathcal{D}_i(n)}^{{\rm N},\mathcal{D}_i}z^{\mathcal{D}_i(n)}}{1+\sum_{m>n}^{N}P_{\mathcal{D}_i(m)}^{{\rm N},\mathcal{D}_i}z^{\mathcal{D}_i(m)}}\right),\, n = 1,2,\ldots,N-1
\end{align}
\end{small}where $z^{\mathcal{D}_i(n)}$ represents the channel gain of source $\mathcal{D}_i(n)$. Under the condition that the other $N-1$ sources' update packets have been correctly decoded at the receiver, the information rate for source $\mathcal{D}_i(N)$ is given by
\begin{small}
\begin{align}
\label{NOMA Action2 X}
R^{{\rm{N}},\mathcal{D}_i}_{\mathcal{D}_i(N)} = \log_2\left(1+P_{\mathcal{D}_i(N)}^{{\rm N},\mathcal{D}_i}z^{\mathcal{D}_i(N)}\right).
\end{align}
\end{small}
Note that an outage occurs for source $\mathcal{D}_i(n)$ if any message from source $\mathcal{D}_i(m)$ with $m<n$ is decoded incorrectly or the message from $\mathcal{D}_i(n)$ is decoded erroneously while all messages from $\mathcal{D}_i(m)$ with $m<n$ are transmitted successfully. Similar to the OMA scenario, let $\overline{R}^{\rm N}$ be the minimum preset rate considering NOMA case which is also same for all the sources. The error transmission probability of source $\mathcal{D}_i(n)$, for all $n=1,2,\ldots,N$, is given by
\begin{small}
\begin{align}
\label{NOMA errorpr}
\epsilon_{\mathcal{D}_i(n)}^{{\rm N},\mathcal{D}_i} &= {\rm Pr}\{ R^{\rm{N},\mathcal{D}_i}_{\mathcal{D}_i(1)} < \overline{R}^{\rm N}\}+(1-{\rm Pr}\{ R^{\rm{N},\mathcal{D}_i}_{\mathcal{D}_i(1)} < \overline{R}^{\rm N}\}){\rm Pr}\{ R^{\rm{N},\mathcal{D}_i}_{\mathcal{D}_i(2)} < \overline{R}^{\rm N}\} \notag \\
&+\ldots+\prod_{m=1}^{n-1}(1-{\rm Pr}\{ R^{\rm{N},\mathcal{D}_i}_{\mathcal{D}_i(m)} <\overline{R}^{\rm N}\}){\rm Pr}\{ R^{\rm{N},\mathcal{D}_i}_{\mathcal{D}_i(n)} < \overline{R}^{\rm N}\} \\
\label{NOMA errorpr_e}
&= \epsilon_{\mathcal{D}_i(1)}^{{\rm N},\mathcal{D}_i}+(1-\epsilon_{\mathcal{D}_i(1)}^{{\rm N},\mathcal{D}_i})\epsilon_{\mathcal{D}_i(2)}^{{\rm N},\mathcal{D}_i}+\ldots+\prod_{m=1}^{n-1}(1-\epsilon_{\mathcal{D}_i(m)}^{{\rm N},\mathcal{D}_i})\epsilon_{\mathcal{D}_i(n)}^{{\rm N},\mathcal{D}_i}.
\end{align}
\end{small}
Likewise, if the decoding order is $\mathcal{D}_j, j \in \{ 1,2,\ldots,N!\}\setminus i $, the corresponding rate and the error transmission probability of each source (i.e., $R^{\rm{N},\mathcal{D}_j}$ and $\epsilon^{{\rm N},\mathcal{D}_{j}}$) can be obtained from a similar slight modification of the above expressions.


To sum up, depending on the current system state, different transmit power levels are adopted. Hence, in both OMA and NOMA cases, the transmit power $P_{n,i}$ of source $n$ belongs to the action space $\mathcal {A} = \{P_{n,1}, P_{n,2}, \ldots, P_{n,K}\}$. Specifically, when $i = K$, $P_{n,i} = 0$, i.e., indicating the idle action. It can be seen that the action $a(m_{n},\Delta_{n}) = P_{n,i}$, channel state gain $z_{i}$ and error transmission probability $\epsilon_{n,i}$ are interrelated.
\subsection{Transition Probabilities}
Assuming that an update packet from source $n$ fails to be decoded by the receiver with the action $a(m_{n},\Delta_{n}) = P_{n,i}$ which occurs with probability $\epsilon_{n,i}$ as defined above, the system enters state $(m_{n}+1,\Delta_{n}+1)$ in the case of $m_{n} < M$. When the transmission round $m_{n}$ reaches the upper bound $M$, the packet is regarded as out of date and a new packet should be generated, which causes the system to enter state $(1,\Delta_{n}+1)$. On the other hand, if the transmission is successful, the state will change to $(1,m_n)$.

Considering that the decoding error probability of a source depends on whether the message decoded earlier is correct or not, the state transitions for $M>2$ sources are more complicated and it is difficult if not intractable to express all state transitions. For instance, if the message for $\mathcal{D}_k(j)$ is decoded incorrectly, messages for all sources $\mathcal{D}_k(i)$ with $i>j$ will be decoded in error definitely, i.e., the state transits from $(m_1,\Delta_1;m_2,\Delta_2;...;m_N,\Delta_N)$ to $(1,m_1;\ldots;1,m_{j-1};m_j+1,\Delta_j+1;...;m_N+1,\Delta_N+1)$ if $m_i<M$, $\forall i\ge j$. Therefore, we mainly concentrate on the two-source scenario. To summarize, we list the transition probabilities of the CMDP problems in a two-source scenario using OMA given by
\begin{small}
\begin{align}
{\rm Pr}\left(m_1+1,\Delta_{1}+1;m_2,\Delta_2 \mid {\mathcal{S}},{\rm P}_{1,i}^{\rm{O}}\right) &= \epsilon_{1,i}^{\rm O},\quad m_1 < M\nonumber \\
{\rm Pr}\left(m_1,\Delta_{1};m_2+1,\Delta_2+1 \mid {\mathcal{S}},{\rm P}_{2,j}^{\rm{O}}\right) &= \epsilon_{2,j}^{\rm O},\quad m_2 < M\nonumber \\
{\rm Pr}\left(1,\Delta_{1}+1;m_2,\Delta_2 \mid {\mathcal{S}},{\rm P}_{1,i}^{\rm{O}}\right) &= \epsilon_{1,i}^{\rm O},\quad m_1 = M\nonumber \\
{\rm Pr}\left(m_1,\Delta_{1};1,\Delta_2+1 \mid {\mathcal{S}},{\rm P}_{2,j}^{\rm{O}}\right) &= \epsilon_{2,j}^{\rm O},\quad m_2 = M\nonumber \\
{\rm Pr}\left(1,m_1;m_2,\Delta_2 \mid {\mathcal{S}},{\rm P}_{1,i}^{\rm{O}}\right) &= 1-\epsilon_{1,i}^{\rm O},\nonumber \\
{\rm Pr}\left(m_1,\Delta_1; 1,m_2 \mid {\mathcal{S}},{\rm P}_{2,j}^{\rm{O}}\right) &= 1-\epsilon_{2,j}^{\rm O},
\end{align}
\end{small}using NOMA:
\begin{small}
\begin{align}
{\rm Pr}\left(m_1+1,\Delta_{1}+1; m_2,\Delta_2 \mid {\mathcal{S}},{\rm P}_{1,i}^{\rm{N}},{\rm P}_{2,j}^{\rm{N}}\right)&= \epsilon_{1,i}^{\rm N}(1-\epsilon_{2,j}^{\rm N}),\quad m_1 < M, \forall m_2 \nonumber \\
{\rm Pr}\left(m_1,\Delta_{1};m_2+1,\Delta_2+1 \mid {\mathcal{S}},{\rm P}_{1,i}^{\rm{N}},{\rm P}_{2,j}^{\rm{N}}\right) &= (1-\epsilon_{1,i}^{\rm N})\epsilon_{2,j}^{\rm N},\quad m_2 < M,\forall m_1 \notag \\
{\rm Pr}\left(1,\Delta_{1}+1;m_2,\Delta_2 \mid {\mathcal{S}},{\rm P}_{1,i}^{\rm{N}},{\rm P}_{2,j}^{\rm{N}}\right)&= \epsilon_{1,i}^{\rm N}(1-\epsilon_{2,j}^{\rm N}),\quad m_1 = M, \forall m_2 \notag \\
{\rm Pr}\left(m_1,\Delta_{1};1,\Delta_2+1 \mid {\mathcal{S}},{\rm P}_{1,i}^{\rm{N}},{\rm P}_{2,j}^{\rm{N}}\right)&= (1-\epsilon_{1,i}^{\rm N})\epsilon_{2,j}^{\rm N},\quad m_2 = M, \forall m_1 \notag \\
{\rm Pr}\left(m_1+1,\Delta_1+1;m_2+1,\Delta_2+1 \mid {\mathcal{S}},{\rm P}_{1,i}^{\rm{N}},{\rm P}_{2,j}^{\rm{N}}\right) &= \epsilon_{1,i}^{\rm N} \epsilon_{2,j}^{\rm N}, \quad m_1 < M, m_2 < M \notag \\
{\rm Pr}\left(m_1+1,\Delta_1+1; 1,\Delta_2+1 \mid {\mathcal{S}},{\rm P}_{1,i}^{\rm{N}},{\rm P}_{2,j}^{\rm{N}}\right)&= \epsilon_{1,i}^{\rm N}\epsilon_{2,j}^{\rm N},\quad m_1 < M, m_2 = M \notag \\
{\rm Pr}\left(1,\Delta_1+1; m_2+1,\Delta_2+1 \mid {\mathcal{S}},{\rm P}_{1,i}^{\rm{N}},{\rm P}_{2,j}^{\rm{N}}\right)&= \epsilon_{1,i}^{\rm N}\epsilon_{2,j}^{\rm N},\quad m_1 = M, m_2 < M \notag \\
{\rm Pr}\left(1,\Delta_1+1; 1,\Delta_2+1 \mid {\mathcal{S}},{\rm P}_{1,i}^{\rm{N}},{\rm P}_{2,j}^{\rm{N}}\right)&= \epsilon_{1,i}^{\rm N}\epsilon_{2,j}^{\rm N},\quad m_1 = M, m_2 = M \notag \\
{\rm Pr}\left(1,m_1; 1,m_2 \mid {\mathcal{S}},{\rm P}_{1,i}^{\rm{N}},{\rm P}_{2,j}^{\rm{N}}\right) &= (1-\epsilon_{1,i}^{\rm N})(1-\epsilon_{2,j}^{\rm N}),
\end{align}
\end{small}and otherwise
\begin{small}
\begin{align}
{\rm Pr}\left(m_1^\prime,\Delta_1^\prime;m_2^\prime,\Delta_2^\prime \mid {\mathcal{S}},{\rm P}_n^{\rm{O}}/{\rm P}_1^{\rm{N}}{\rm P}_2^{\rm{N}}\right) &= 0,
\end{align}
\end{small}where ${\mathcal{S}}$ represents the current state $(m_1,\Delta_1;m_2,\Delta_2)$. 

Note that our state space is countably infinite, since the value of AoI can be arbitrarily large. In practice, however, we can use a large but finite space to approximate the countably infinite state space by setting an upper bound on the age ( which will be denoted by $\Delta_{max}$). In such way, whenever the AoI exceeds $\Delta_{max}$, we set it to be 1. Clearly, when $\Delta_{max}$ is close to infinity, the optimal policy for the finite state space will converge to that of the original problem.

\subsection{Rewards, Costs and Problem Formulation}
For source $n$, let us define the decision $\mu_{n,t}(s)$ as a function that maps the system state $s_t =(m_{1,t},\Delta_{1,t};m_{2,t},\Delta_{2,t};\ldots;m_{N,t},\Delta_{N,t})$ to the action $a(m_n,\Delta_n)$ to be taken in time block $t$. And the policy $\pmb{\mu_n} = \{\mu_{n,1},\mu_{n,2},\ldots\}$ is called stationary if the actions are independent of time slot $t$. Therefore, we denote $s_t^{\mu_n}$ as sequences of states induced by policy $\pmb{\mu_n}$. For a stationary policy $\pmb{\mu_n}$, we define the \textit{reward} function of the CMDP problems (i.e. the long term weighted sum average AoI) in both OMA and NOMA environments as
\begin{small}
\begin{align}
\label{longtermaoi}
\widetilde{\Delta}(s,\mu) =   (1-\lambda)\sum_{n=1}^{N} \sum_{t=1}^{\infty}\lambda^{t-1}\mathbb{E}[\omega_n (\Delta_{n,t}^{\mu}+\frac{1}{2})],
\end{align}
\end{small}where $\lambda \in (0,1)$ is a discount factor, $\omega_n > 0$ is the weight coefficient representing the priority of source $n$ and $\mathbb{E}[\cdot]$ is the expectation operator which are taken over the policy $\pmb{\mu_n}$. Then the \textit{cost} function (i.e. the long term average power consumption) for source $n$ on the OMA scheme is given by
\begin{small}
\begin{align}
\label{longtermPoma}
\widehat{\mathrm{P}}_{n}^{\mathrm{O}}(s,\mu) = (1-\lambda)\sum_{t=1}^{\infty}\lambda^{t-1}\mathbb{E}[\mathrm{P}_{n,t}^{\mathrm{O},\mu}],
\end{align}
\end{small}and the \textit{cost} function for source $n$ when NOMA is employed can be written as
\begin{small}
\begin{align}
\label{longtermPnoma}
\widehat{\mathrm{P}}_{n}^{\mathrm{N}}(s,\mu) = (1-\lambda)\sum_{t=1}^{\infty}\lambda^{t-1}\mathbb{E}[\mathrm{P}_{n,t}^{\mathrm{N},\mu}].
\end{align}
\end{small}Note that in (\ref{longtermaoi}), (\ref{longtermPoma}) and (\ref{longtermPnoma}), as $\lambda \rightarrow 1$, the infinite sum of the discounted rewards and costs will converge to their corresponding expected average rewards and costs \cite{M. H. Ngo}. Then the formulated CMDP problems on both OMA and NOMA schemes can be expressed respectively as

\emph{Problem 1 (CMDP optimization problem in OMA):}
\begin{small}
\begin{align}
\label{pro:oma}
\min_{\mu} \,\, &\widetilde{\Delta}(s,\mu)  \\
\text{s.t.}\,\,&\widehat{\mathrm{P}}_{n}^{\mathrm{O}}(s,\mu) \leq \frac{\overline{P}_n}{\rho_n},\quad \forall n \nonumber
\end{align}
\end{small}
\emph{Problem 2 (CMDP optimization problem in NOMA):}
\begin{small}
\begin{align}
\label{pro:noma}
\min_{\mu} \,\, &\widetilde{\Delta}(s,\mu) \\
\text{s.t.}\,\,&\widehat{\mathrm{P}}_{n}^{\mathrm{N}}(s,\mu) \leq \overline{P}_n,\quad \forall n \nonumber
\end{align}
\end{small}where $\overline{P}_n$ is the individual constraint on the average power of source $n$.

And please note that the error transmission probability $\epsilon_{n,i}$ is only available when CDI is given. However, when environment is not known as a priori, such information is not accessible to us and the CMDP optimization problems mentioned above are no longer applicable, as will be discussed in Section V.
\section{Optimal Policy with CDI at Transmitters }
In this section, we adopt the Lagrangian relaxation method to solve the problems mentioned above. In this way, a CMDP problem can be converted into a equivalent unconstrained MDP problem by introducing the a non-negative multiplier $\beta$ and \emph{Lagrangizan reward} for source $n$ is defined as
\begin{small}
\begin{align}
\label{lag:reward}
r_n(s,\mu_n;\beta_n) &=  \omega_n(\Delta_n(s,\mu_n)+\frac{1}{2})+\beta_{n} p_{n}(s,\mu_n),
\end{align}
\end{small}where $p_{n}(s,\mu)$ denotes the power consumption in state $s$ with policy $\mu_n$ for the OMA or NOMA scheme. Since expression (\ref{lag:reward}) is valid for both transmission modes, we omit the superscripts from the notation for simplicity here. And then, accordingly, there exists a \emph{value function} $V_{n,\beta_n}(s)$ satisfying
\begin{small}
\begin{align}
\label{lag:valuefunction}
V_{n,\beta_n}(s) = \min_{a\in \mathcal{A}}\left\{(1-\lambda)r_n + \lambda\sum_{s'\in \mathrm{S}}\mathrm{Pr}\{s' \mid s,a\}V_{n,\beta_n}(s') \right\},
\end{align}
\end{small}called the Bellman optimality equations for source $n$ in all states $s \in \mathcal{S}$, where $s'$ denotes the next state obtained from state $s$ after taking action $a$. Hence, the policy for source $n$ can be indicated as follows:
\begin{align}
\label{lag:policy}
\mu_{n}^{\beta_n}(s) \triangleq \arg\min_{a\in \mathcal{A}}V_{n,\beta_n}(s).
\end{align}

A popular and effective approach, called Value Iteration Algorithm (VIA), is adopted here to solve (\ref{lag:valuefunction}) for any given Lagrangian multiplier $\beta$. With the any initialization of $V_{n,\beta_n}^{0}(s), \forall s$, $V_{n,\beta_n}^{k+1}(s)$ will be updated in each iteration satisfying
\begin{small}
\begin{align}
\label{lag:valueiteration}
V_{n,\beta_n}^{k+1}(s) = \min_{a\in \mathcal{A}}\left\{(1-\lambda)r_n + \lambda\sum_{s'\in \mathrm{S}}\mathrm{Pr}\{s' \mid s,a\}V_{n,\beta_n}^{k}(s') \right\},
\end{align}
\end{small}until convergence.

Based on \cite{L. I. Sennott}-\cite{M. H. Ngo}, there exists optimal stationary policies for the formulated CMDP problems, which are also optimal for the corresponding unconstrained problems considered in (\ref{lag:reward}) for some $\beta_n = \beta_n^*$. To simplify the notation, the dependence on $n$ is suppressed in the following. For arbitrary source, this optimal policy $\mu^*$ is a probabilistic mixture of two deterministic policies $\mu_{\beta^+}$ and $\mu_{\beta^-}$. To be more precise, there exists $\xi \in [0,1]$ such that the optimal policy $\mu^*(s)$, in any state $s$, selects the action $\mu_{\beta^-}$ with probability $\xi$ and $\mu_{\beta^+}$ with probability $1-\xi$. Let $\widehat{\mathrm{P}}_\beta$ be the average power consumption associated with the policy $\mu_\beta(s)$. According to the fact that $\widehat{\mathrm{P}}_\beta$ is monotonically decreasing when $\beta$ increases, by bisection method, we can find the values of $\beta^-$ and $\beta^+$ which are slightly smaller and larger than $\beta^*$, respectively. Therefore, according to the given $\beta^-$ and $\beta^+$, the steady state distribution $\pi^-$ and $\pi^+$ can be calculated and we can accordingly obtain the average AoI $\overline{\Delta}$ and average power consumption $\widehat{\mathrm{P}}$ under these two policies as follows:
\begin{small}
\begin{align}
\label{lag:AOI}
\overline{\Delta}_{\beta^\pm} &=\sum_s\Delta(s)\pi^{\beta^\pm}(s), \\
\label{lag:P}
\widehat{\mathrm{P}}_{\beta^\pm} &= \sum_s\mu^{\beta^\pm}(s)\pi^{\beta^\pm}(s).
\end{align}
\end{small}Then the corresponding mixing coefficients $\xi$ in OMA and NOMA environments can be derived through
\begin{small}
\begin{align}
\label{lag:xiO}
\xi^\mathrm{O} \widehat{\mathrm{P}}_{\beta^-}^\mathrm{O} + (1-\xi^\mathrm{O})\widehat{\mathrm{P}}_{\beta^+}^\mathrm{O} &= \overline{\mathrm{P}}/\rho,\\
\label{lag:xiN}
\xi^\mathrm{N} \widehat{\mathrm{P}}_{\beta^-}^{\mathrm{N}} + (1-\xi^\mathrm{N})\widehat{\mathrm{P}}_{\beta^+}^\mathrm{N} &= \overline{\mathrm{P}},
\end{align}
\end{small}respectively. To sum up, the pseudo code of the VIA in OMA is given in Algorithm 1 and that in NOMA is provided in Algorithm 2.
\begin{algorithm}[htb]
\centering
\caption{Proposed Optimal Policy in OMA}
\label{alg:A1}
\begin{algorithmic}[1]
\STATE\textbf{Input:} $ \overline{\mathrm{P}}, K, M, \Delta_{max}, R ,N$;
\STATE \textbf{Initialization:} $\beta_n^- = 0, \beta_n^+, \mathrm{V}_{n,\beta_n}^{0} = 0$, where $n\in[N]$, and specify $\Gamma_{\beta},\Gamma_{\mathrm{V}}$;
\WHILE{$\max_n |\beta_n^+ - \beta_n^-| > \Gamma_{\beta}$}
\STATE Let $\beta_n = \beta_n^i = \left(\beta_n^- + \beta_n^+\right)/2$, $i = i+1$, $k = 0$;
\WHILE{$\max_s |\mathrm{V}_{\beta}^{k+1}(s) - \mathrm{V}_{\beta}^{k}(s)|  > \Gamma_{\mathrm{V}}$}
\FOR{$n = 1$ \TO $N$}
\STATE $V_{n,\beta_n}^{k+1}(s) =\min_{a\in \mathcal{A}}\left\{(1-\lambda)r_n +\lambda\sum_{s'\in \mathrm{S}}\mathrm{Pr}\{s' \mid s,a\}V_{n,\beta_n}^{k}(s') \right\},\forall s;$
\ENDFOR
\STATE $V_{\beta}^{k+1}(s) = \sum_{n=1}^N V_{n,\beta_n}^{k+1}(s)$;
\STATE $k = k+1;$
\ENDWHILE
\STATE Derive the corresponding policy $\mu_{n}^{\beta_n}(s)$ from (\ref{lag:policy}) and optimal values of $\rho_n$;
\STATE Compute the corresponding steady state distribution $\pi_n^{\beta_n}(s)$ and \emph{cost} in OMA $\widehat{\mathrm{P}}_{n,\beta_n}^\mathrm{O} = \sum_s\mu_{n}^{\beta_n}(s)\pi_n^{\beta_n}(s)$;
\IF{$\widehat{\mathrm{P}}_{n,\beta_n}^\mathrm{O} > \frac{\overline{P}_n}{\rho_n}$}
\STATE $\beta_n^- = \beta_n$;
\ELSE
\STATE $\beta_n^+ = \beta_n$;
\ENDIF
\ENDWHILE
\STATE Compute $\overline{\Delta}_{n,\beta_n^-}, \overline{\Delta}_{n,\beta_n^+}, \widehat{\mathrm{P}}_{n,\beta_n^-}^\mathrm{O}, \widehat{\mathrm{P}}_{n,\beta_n^+}^\mathrm{O}$ and $\xi^\mathrm{O}_n$ using (\ref{lag:AOI})-(\ref{lag:xiO});
\STATE Compute $\overline{\Delta}_{n} = \xi^\mathrm{O}_n\overline{\Delta}_{n,\beta_n^-} + (1-\xi^\mathrm{O}_n)\overline{\Delta}_{n,\beta_n^+}$;
\STATE The optimal weighted sum average AoI is derived:
\STATE $\widetilde{\Delta} =\sum_{n=1}^N \omega_n\overline{\Delta}_{n} $.
\end{algorithmic}
\end{algorithm}

\begin{algorithm}[htb]
\centering
\caption{Proposed Optimal Policy in NOMA}
\label{alg:A2}
\begin{algorithmic}[1]
\STATE\textbf{Input:} $ \overline{\mathrm{P}}, K, M, \Delta_{max}, R, N$;
\STATE \textbf{Initialization:} $\beta_n^- = 0, \beta_n^+, \mathrm{V}_{n,\beta_n}^{0} = 0$, where $n\in[N]$, and specify $\Gamma_{\beta},\Gamma_{\mathrm{V}}$;
\WHILE{$\max_n |\beta_n^+ - \beta_n^-| > \Gamma_{\beta}$}
\STATE Let $\beta_n = \beta_n^i = \left(\beta_n^- + \beta_n^+\right)/2$, $i = i+1$, $k = 0$;
\WHILE{$\max_s |\mathrm{V}_{\beta}^{k+1}(s) - \mathrm{V}_{\beta}^{k}(s)|  > \Gamma_{\mathrm{V}}$}
\STATE /*\quad Select the optimal decoding order $\mathcal{D}^*$ that minimizes the \emph{value function}\,\,\,\,*/
\FOR{$n = 1$ \TO $N$}
\STATE $V_{\mathcal{D}^*(n),\beta_n}^{k+1}(s) =\min_{a\in \mathcal{A}}\left\{(1-\lambda)r_n + \lambda \mathbb{E}[V_{\mathcal{D}^*(n),\beta_n}^{k,}(s)]\right\},\,\, \mathcal{D}^* \in \pmb{\mathcal{D}};$
\ENDFOR
\STATE $V_{\beta}^{k+1}(s) = \sum_{n=1}^N V_{\mathcal{D}^*(n),\beta_n}^{k+1}(s)$;
\STATE $k = k+1;$
\ENDWHILE
\STATE Derive the corresponding policy $\mu_{n}^{\beta_n}(s)$ from (\ref{lag:policy});
\STATE Compute the steady state distribution $\pi_n^{\beta_n}(s)$ and \emph{cost} in NOMA $\widehat{\mathrm{P}}_{n,\beta_n}^\mathrm{N} = \sum_s\mu_{n}^{\beta_n}(s)\pi_n^{\beta_n}(s);$
\IF{$\widehat{\mathrm{P}}_{n,\beta_n}^\mathrm{N} > \overline{P}_n$}
\STATE $\beta_n^- = \beta_n$;
\ELSE
\STATE $\beta_n^+ = \beta_n$;
\ENDIF
\ENDWHILE
\STATE Compute $\overline{\Delta}_{n,\beta_n^-}, \overline{\Delta}_{n,\beta_n^+}, \widehat{\mathrm{P}}_{n,\beta_n^-}^\mathrm{N}, \widehat{\mathrm{P}}_{n,\beta_n^+}^\mathrm{N}$ and $\xi^\mathrm{N}_n$ using (\ref{lag:AOI}), (\ref{lag:P}) and (\ref{lag:xiN});
\STATE Compute $\overline{\Delta}_{n} = \xi^\mathrm{N}_n\overline{\Delta}_{n,\beta_n^-} + (1-\xi^\mathrm{N}_n)\overline{\Delta}_{n,\beta_n^+}$;
\STATE The optimal weight sum average AoI is derived:
\STATE $\widetilde{\Delta} =\sum_{n=1}^N \omega_n\overline{\Delta}_{n} $.
\end{algorithmic}
\end{algorithm}

\section{Reinforcement Learning Algorithm in an Unknown Environment}
In the above section, it is assumed that CDI and error transmission probability are known in advance. However, in most practical scenarios, CDI is not available or may continuously change over time. So, in this section, we assume the sources do not have a priori information about CDI and have to learn it. As a result, we adopt online reinforcement learning (RL) approaches to minimize weighted sum AoI in OMA and NOMA without degrading the performance significantly compared with VIA proposed above. In RL, an agent constantly interacts with the environment to learn a good scheduling policy without prior information. Then, at each iteration, the agent observes state $\mathcal{S}^i$ and takes an action $a^i$. After performing the action $a^i$, the current state will transmit to $\mathcal{S}^{i+1}$, and the agent will receive corresponding \emph{reward} $r^{i+1}$ and the \emph{cost} of this step $p^{i}$. Above, $i$ is the index of iterations.

Q-learning is one of the most popular reinforcement learning algorithms, in which the agent obtains corresponding reward by interacting with the environment and gradually fills the Q-table by calculating the discounted accumulative rewards termed the state-action function, i.e., $Q(\mathcal{S},a)$. At each iteration, the Q-values are updated and stored in the Q-table again. After learning sufficiently many number of iterations or enough time, each Q-value will converge to a certain fixed value and the agent will find the best policy for our problems based on the minimum non-zero value in the Q-table. The update process of the Q-function is shown as follows:
\begin{small}
\begin{align}
\label{Q-func}
Q_n(\mathcal{S}^i,a_n^i)&\leftarrow Q_n(\mathcal{S}^i,a_n^i) +\alpha_n^i[r_n^{i+1}+\lambda\min_{a_n^{\prime}}Q_n(\mathcal{S}^{i+1},a_n^{\prime}) -Q_n(\mathcal{S}^{i},a_n^i)].
\end{align}
\end{small}
Nevertheless, there always exist some problems in Q-learning, such as curse of dimensionality and difficulty in convergence \cite{J. Moody}, \cite{L. Baird}. What's more, as discussed in section III, there are $K$ possible power values for the agent to choose from in this work. So how to balance the exploration and exploitation in learning process is extremely important, which determines how fast the algorithm converges and how long this learning process takes. Otherwise, adopting inappropriate policies will undermine our learning process. Here we adopt Q-learning with $\epsilon-$greedy exploration algorithm to find the optimal policy for both OMA and NOMA schemes. This means that, at the beginning of each training session, the agent selects a random action with some probability $\epsilon$ and execute a greedy policy (i.e., selects a historically optimal action according to Q-table) with probability $1-\epsilon$, as below:
\begin{small}
\begin{equation}
\label{epsilon-greedy}
\mu(s) = \mathrm{Pr}(a|s)=\left\{
\begin{aligned}
&\frac{\epsilon}{K}+1-\epsilon \quad &\mathrm{if}\, a^* &= \arg\min_{a\in \mathcal{A}}Q(\mathcal{S},a)\\
&\frac{\epsilon}{K} &\quad else
\end{aligned}
\right
.
\end{equation}
\end{small}where $\epsilon$ is the exploration probability which ususally diminishes with the iterative process of the algorithm and eventually tends to 0. In this way, during the early iterative period, we encourage exploration, whereas after a sufficient number of iterations, as we have enough exploration, the policy asymptotically becomes conservative and greedy,  so that the algorithm can converge stably.

For the purpose of comparing AoI performance with known environment, for the source $n$, we employ the same transmit power set (also known as action space) $\mathcal {A} = \{P_{n,1}, P_{n,2}, \ldots, P_{n,K}\}$ discussed in section III.B and select action according to $\epsilon-$greedy exploration algorithm at the beginning. To speed up the convergence of Q-learning, we normalize the transmit power set using the average value of the power set and the average transmitted power constraint, i.e., $\| \mathcal{A}^\mathrm{O }\| = \left\{\frac{KP_{n,k}}{\sum_{k=1}^{K}P_{n,k}}\frac{\overline{P}_n}{\rho_n}\right\}$ and $\| \mathcal{A}^\mathrm{N} \| =\left\{\frac{KP_{n,k}}{\sum_{k=1}^{K}P_{n,k}}\overline{P}_n\right\}$, $k=1,2,\ldots,K$, for OMA and NOMA, respectively. Numerical simulations show a considerable saving in training time by normalization without any significant change in performance. The details of Q-learning in OMA and NOMA are given in Algorithm 3 and Algorithm 4, respectively.
\begin{algorithm}[htb]
\centering
\caption{Q-Learning in OMA}
\label{alg:QL1}
\begin{algorithmic}[1]
\STATE\textbf{Input:} $ \overline{\mathrm{P}}, K, M, \Delta_{max}, R, \lambda, \beta, I, \|\mathcal{A}^\mathrm{O}\|, N$;
\STATE \textbf{Initialization:} $ Q^{M\ast\Delta_{max}\ast K}_n \leftarrow \pmb{0},$ where $n\in[N],i \leftarrow 0 , \mathcal{S}^0=(s_1^0;s_2^0;\ldots;s_N^0)$;
\FOR{$n = 1$ \TO $N$}
\WHILE{$i \leq I$}
\STATE $a_n^i\leftarrow$ choose an action from $\mathcal{A}$ based on \texttt{$\epsilon$-greedy} algorithm for $\mathcal{S}^i$;
\STATE Observe next state $\mathcal{S}^{i+1}$, and calculate corresponding \textit{reward} function $r_n^{i+1}$;
\STATE Get average age $\overline{\Delta}_n^{i}$ and \textit{cost} function $p_n^{i}$;
\STATE\textbf{Update}
\STATE $\alpha_n^i = \frac{1}{\sqrt{i}}$; \quad\quad  /*\texttt{\quad update parameter \quad}\quad*/
\STATE $Q_n(\mathcal{S}^i,a_n^i)\leftarrow Q_n(\mathcal{S}^i,a_n^i)+\alpha_n^i[r_n^{i+1} +\lambda\min_{a_n^{\prime}}Q_n(\mathcal{S}^{i+1},a_n^{\prime}) -Q_n(\mathcal{S}^{i},a_n^i)]$;
\STATE $i \leftarrow i+1;$  /*\,\, \texttt{increase the iteration }\,\,*/
\ENDWHILE
\ENDFOR
\STATE /*\,\,\texttt{\,\,\, Derive weighted sum average AoI \,\,\,}\,\,\,*/
\STATE $\widetilde{\Delta} =\sum_{n=1}^N \omega_n\overline{\Delta}_n^{I}$.
\end{algorithmic}
\end{algorithm}

\begin{algorithm}[htb]
\centering
\caption{Q-Learning in NOMA}
\label{alg:QL2}
\begin{algorithmic}[1]
\STATE\textbf{Input:} $ \overline{\mathrm{P}}, K, M, \Delta_{max}, R, \lambda, \beta, I, \|\mathcal{A}^\mathrm{N}\|$;
\STATE \textbf{Initialization:} $ Q^{M\ast\Delta_{max}\ast K}_n \leftarrow \pmb{0},$ where $n\in[N],i \leftarrow 0 , \mathcal{S}^0=(s_1^0;s_2^0;\ldots;s_N^0)$ and specify $\Gamma_{\Delta}$;
\WHILE{$|\widetilde{\Delta}-\widetilde{\Delta}^{\prime}| < \Gamma_{\Delta}$}
\WHILE{$i \leq I$}
\FOR {$k = 1$ \TO $N!$}
\STATE Set decoding order $\mathcal{D} = \mathcal{D}_k:$
\STATE Do Algorithm 3 (5-7) to get $a_{\mathcal{D}_k(n)}^i, r_{\mathcal{D}_k(n)}^{i+1}, \overline{\Delta}_{\mathcal{D}_k(n)}^{i}$ and $p_{\mathcal{D}_k(n)}^{i}, n\in [N] $;
\ENDFOR
\STATE\textbf{Find} optimal decoding order $\mathcal{D}^*$ according to the \emph{reward} function:
\STATE $a_n^i \leftarrow a_{\mathcal{D}^*(n)}^i , \overline{\Delta}_n^{i} \leftarrow \overline{\Delta}_{\mathcal{D}^*(n)}^{i}$ and $p_n^{i} \leftarrow p_{\mathcal{D}^*(n)}^{i}, n \in [N];$
\STATE\textbf{Update}
\STATE $\alpha_1^i = \alpha_2^i = \ldots = \alpha_N^i = \frac{1}{\sqrt{i}}$; \quad  /*\texttt{\quad update parameter \quad}*/
\FOR {$n= 1$ \TO $N$}
\STATE $Q_n(\mathcal{S}^i,a_n^i)\leftarrow Q_n(\mathcal{S}^i,a_n^i)+\alpha_n^i[r_n^{i+1} +\lambda\min_{a_n^{\prime}}Q_n(\mathcal{S}^{i+1},a_n^{\prime}) -Q_n(\mathcal{S}^{i},a_n^i)]$;
\ENDFOR
\STATE $i \leftarrow i+1;$  /*\,\, \texttt{increase the iteration }\,\,*/
\ENDWHILE

\STATE /*\,\texttt{\,\, Derive weighted sum average AoI \,\,}*/
\STATE $\widetilde{\Delta} =\sum_{n=1}^N \omega_n\overline{\Delta}_n^{I}$.

\ENDWHILE
\end{algorithmic}
\end{algorithm}

In Algorithm 3 and Algorithm 4, $\alpha_n^i$ is the update parameter (or learning rate equivalently) in the $i$-th iteration for source $n$. However, in practice, it is hard to find the optimal Lagrange multiplier $\beta^*$ which satisfies $\widehat{\mathrm{P}}_{\beta}^\mathrm{O}\thickapprox \frac{\overline{P}}{\rho}$ or $\widehat{\mathrm{P}}_{\beta}^\mathrm{N} \thickapprox \overline{P}$. As a result, we have the following heuristic as in \cite{E. T. Ceran1}: To find the desired non-negative optimal $\beta^*$, we run the iterative algorithms via stochastic gradient descent with an initialized parameter $\beta^0$, which are given by
\begin{small}
\begin{align}
\label{update_beta_OMA}
\beta^{i+1}&= \max\left\{0,\beta^{i}+\zeta^i(\widehat{\mathrm{P}}_{\beta^i}^\mathrm{O}- \frac{\overline{P}}{\rho})\right\}, \\
\label{update_beta_NOMA}
\beta^{i+1}&= \max\left\{0,\beta^{i}+\zeta^i(\widehat{\mathrm{P}}_{\beta^i}^\mathrm{N} - \overline{P})\right\},
\end{align}
\end{small}for OMA and NOMA, respectively. $\zeta^i$ is a positive and deceasing sequence which satisfies the following conditions:
\begin{small}
\begin{align}
\sum_i^\infty \zeta^i = \infty, \qquad \sum_i^\infty (\zeta^i)^2 = 0.
\end{align}
\end{small}
That is to say, we update the Lagrange multipliers based on the empirical power consumption in order to solve the proposed AoI optimization problems. In each iteration, we keep track of a value $\beta$ which drives the transmission cost close to power constraint and then derive the optimal policy. Please note that, in Algorithm 4, we must ensure that the gap between the average weighted sum AoI obtained from two consecutive training processes is smaller than a given threshold (i.e., $|\widetilde{\Delta}-\widetilde{\Delta}^{\prime}| < \Gamma_{\Delta}$) to guarantee the accuracy and reliability of the algorithm. This is due to the fact that NOMA requires to update the Q-values of both sources and their corresponding power consumption simultaneously, which makes it quite problematic to achieve convergence of the algorithm to obtain the desired results. By contrast, we have no such issues to worry about in OMA's Q-learning, since in OMA the multiple sources are learning to update separately.
\section{Numerical Results}
In this section, we focus mainly on two-source case and present some numerical results to validate our theoretical analyses. Unless otherwise specified, we assume that $M=4, \Delta_\mathrm{max}=100$ to approximate the countably infinite state space and that $\overline{R}^{\rm{O}} = \overline{R}^{\rm{N}} = R = 1.7$ bits (or bps/Hz equivalently), $\lambda=0.99, K = 128$ and $\omega_1 = \omega_2 = 1$ in the following. Rayleigh fading channels with unit mean are considered here, and in the off-line value iteration algorithms, we assume the channel state probabilities are $\psi_0=\psi_1=\ldots=\psi_{K-1}=\frac{1}{K}$. In reinforcement learning algorithms, $I = 10000$ as the maximum number of iterations for each episode.
\subsection{Performance in a Known Environment}
\begin{figure}
    \centering
    \includegraphics[width=0.5\textwidth]{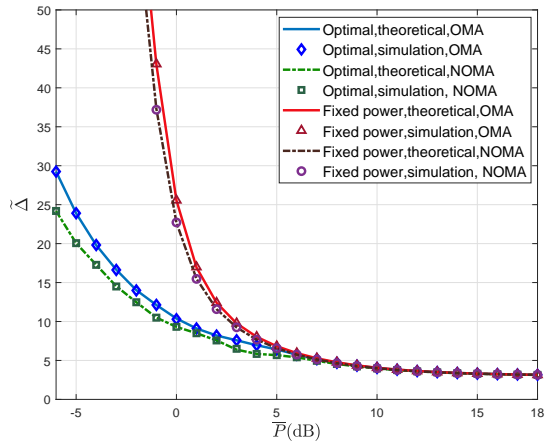}
    \caption{The weighted sum average AoI as a function of $\overline{P}$.}
    \label{fig:OMANOMA}
\end{figure}

\begin{figure}
    \centering
    \includegraphics[width=0.5\textwidth]{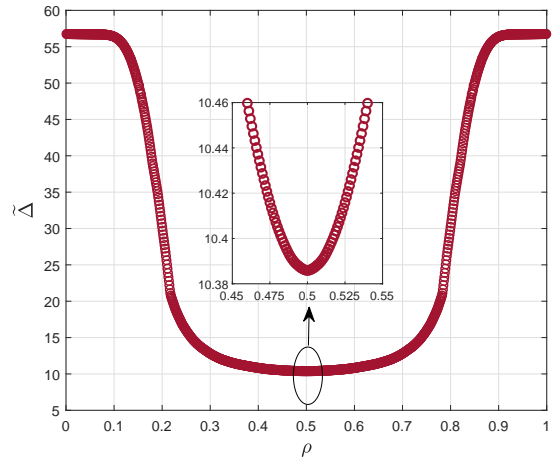}
    \caption{The weighted sum average AoI as a function of $\rho$. $\overline{P}$ = 0 dB.}
    \label{fig:OMANOMA2}
\end{figure}

First, in Fig. \ref{fig:OMANOMA}, we plot the weighted sum average AoI on OMA and NOMA schemes as a function of $\overline{P}$, where $\overline{P} = \overline{P}_1 = \overline{P}_2 $ such that the subscript $n$ is neglected for simplification. ``Fixed power" refers to the policy that sets the transmit power to a fixed level, and ``theoretical'' stands for the statistical values derived from the algorithms directly. Moreover, we simulate the corresponding policies over $10^6$ time slots and then take the average age over time for simulation. In this way we obtain the ``simulation" curves accordingly. What stands out in this figure is that the theoretical calculation results match well with the simulation results on both OMA and NOMA schemes, which confirms the validity of the proposed optimal policy. Here we can see that NOMA always outperforms OMA when $R=$ 1.7 bps/Hz, which is reflected in a lower AoI. Compared with the ``fixed power'' policy, the proposed optimal policy reduces age significantly, especially in the low-power regime. As $\overline{P}$ increases successively, both policies further converge to a similar performance.

\begin{figure}
    \centering
    \subfigure[The derived optimal $\rho$ as a function of $\alpha$.]{
    \includegraphics[width=0.47\textwidth]{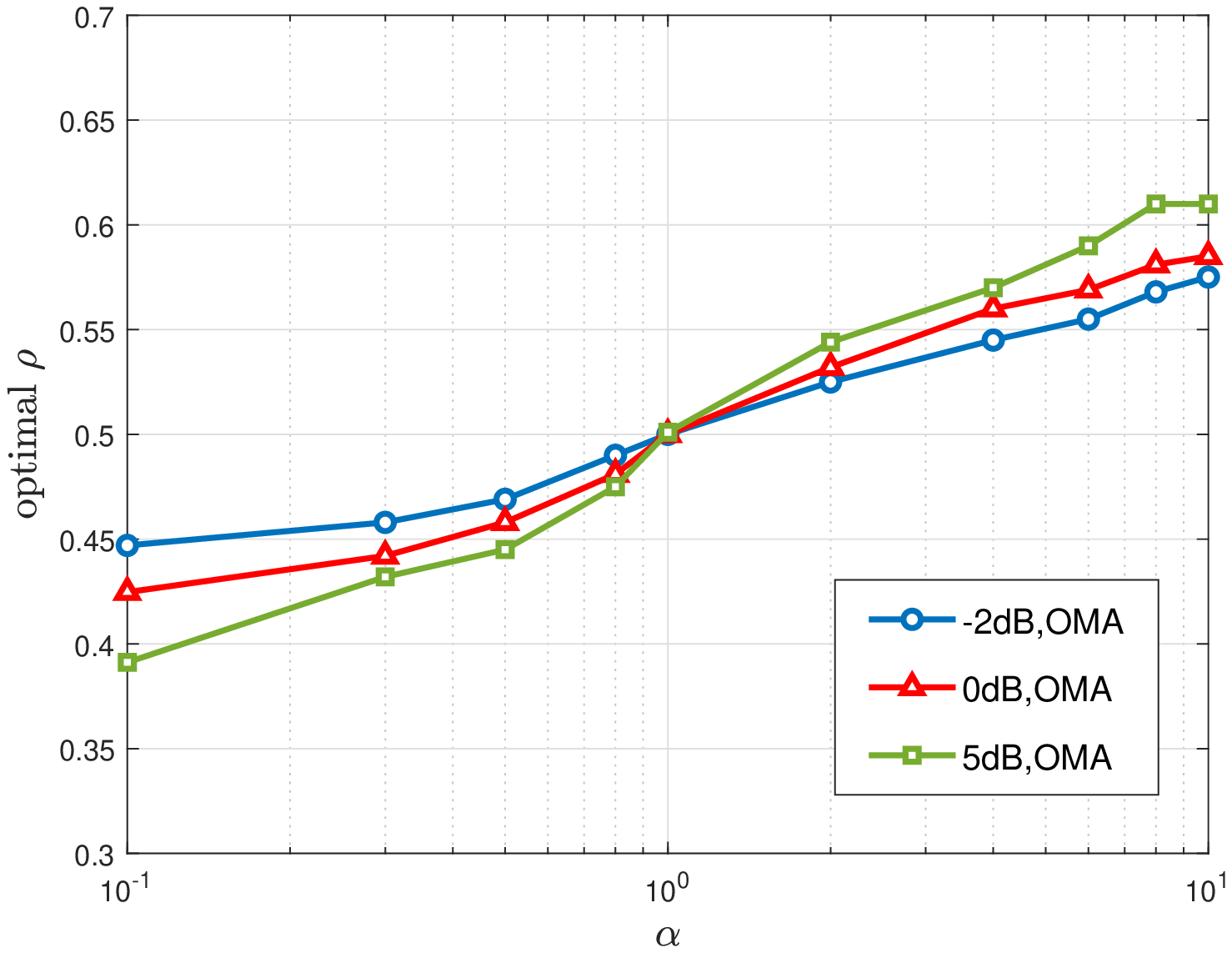}
    \label{fig:optiamlrho_alpha}
    }
    \subfigure[The weighted sum average AoI as a function of $\alpha$.]{
    \includegraphics[width=0.45\textwidth]{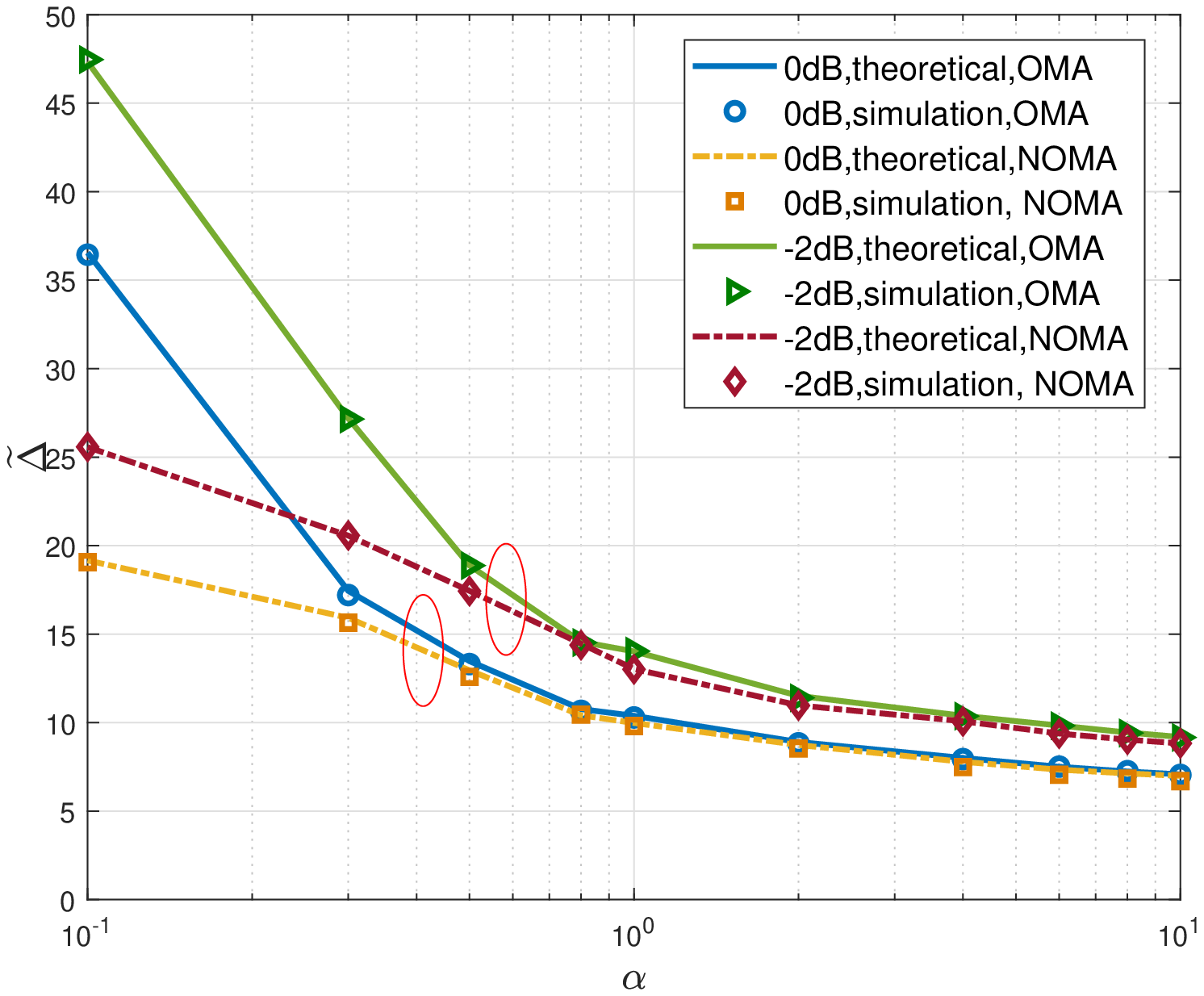}
    \label{fig:OMANOMA3}
    }
    \caption{Performance when $\overline{P}_2$ changes. }
    \label{fig:Performance in P2}
\end{figure}

Next, based on the Algorithm 1, we put the focus on finding the optimal $\rho$ (in two-source case, we set $\rho_1=\rho$ and $\rho_2 = 1-\rho$) that minimizes the AoI. In Fig. \ref{fig:OMANOMA2}, we plot the weighted sum average AoI as a function of $\rho$ with the assumptions of $\overline{P} = 0$ dB. Generally speaking, the average AoI reaches the minimum when $\rho$ is around 0.5 (i.e. the update packet sent by source 1 occupies half of one time slot) since the initial conditions of the two sources are set to be the same. However, when the average power values change, the value of the optimal $\rho$ will change accordingly, e.g., as shown in Fig. \ref{fig:optiamlrho_alpha}.

We first fix the power constraint $\overline{P}_1$ for source 1 and then adjust power constraint $\overline{P}_2$ for source 2 by the proportion $\alpha$ (i.e., $\overline{P}_1=\overline{P}, \overline{P}_2 = \alpha\overline{P}$). In Fig. \ref{fig:optiamlrho_alpha}, we plot the derived optimal $\rho$ as the coefficient $\alpha$ varies from 0.1 to 10 under different power constraints, indicating a proportional reduction or amplification of $\overline{P}_2$. Obviously, the optimal $\rho$ increases with $\alpha$ in all cases, indicating that source 1 can occupy more time fraction for status update if the average power of source 2 becomes larger. Moreover, we can see from the figure that the difference of the transmission time between the sources becomes smaller if the average power levels are low. This is due to the fact that in the low-power regime, more time is needed to successfully send update packets for each source, and hence a wise choice would be compromise between the sources.

Likewise, in Fig. \ref{fig:OMANOMA3}, we investigate the weighted sum average AoI as a function of $\alpha$ for different multiple access schemes when $\overline{P} = -2$ dB and $0$ dB. We notice that the weighted sum average AoI will increase significantly when $\alpha$ is less than 1 especially in OMA case. In other words, reducing one source's power constraint will obviously lead to a poor performance. On the other hand, when $\alpha > 1$, age of information does not decrease dramatically with increases in $\alpha$ due to the fact that power constraint is large enough that a larger $\overline{P}_2$ will not lead to a obvious performance improvement, especially for a relatively large power constraint as can be seen from the comparison of $\overline{P} = 0$ dB and $\overline{P} = -2$ dB in the figure.
\begin{figure}
    \centering
    \includegraphics[width=0.5\textwidth]{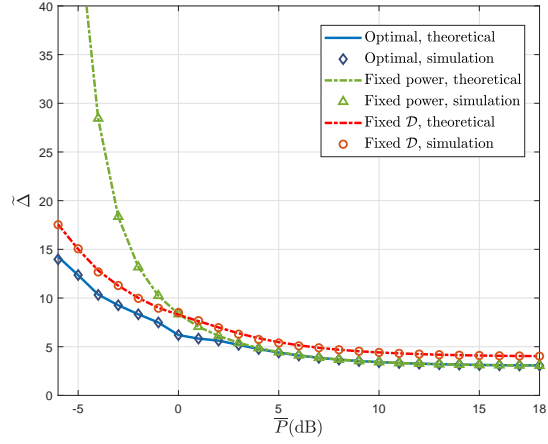}
    \caption{ AoI performance under different schemes in NOMA when $R =$ 1 bps/Hz.}
    \label{fig:OMANOMA4}
\end{figure}

Fig. \ref{fig:OMANOMA4} shows the weighted sum average AoI performance under different schemes in NOMA. Here we assume R = 1 bps/Hz. ``Fixed $\mathcal{D}$'' is the policy that arranges to send the update packet with a fixed decoding order in NOMA, which is a newly proposed scheme to investigate the impact of decoding order $\mathcal{D}$ on AoI performance. This result is somewhat interesting in which the gap between blue curve and red curve does not change very noticeably. This is because the fixed decoding order is a rather independent and slight point leading to performance deterioration, which indicates that it does not change as drastically as ``Fixed power''. On the other hand, this leads to a failure to achieve optimal performance until convergence.

\begin{figure}
    \centering
    \subfigure[Optimal decoding order $\mathcal{D}$.]{
    \includegraphics[width=0.47\textwidth]{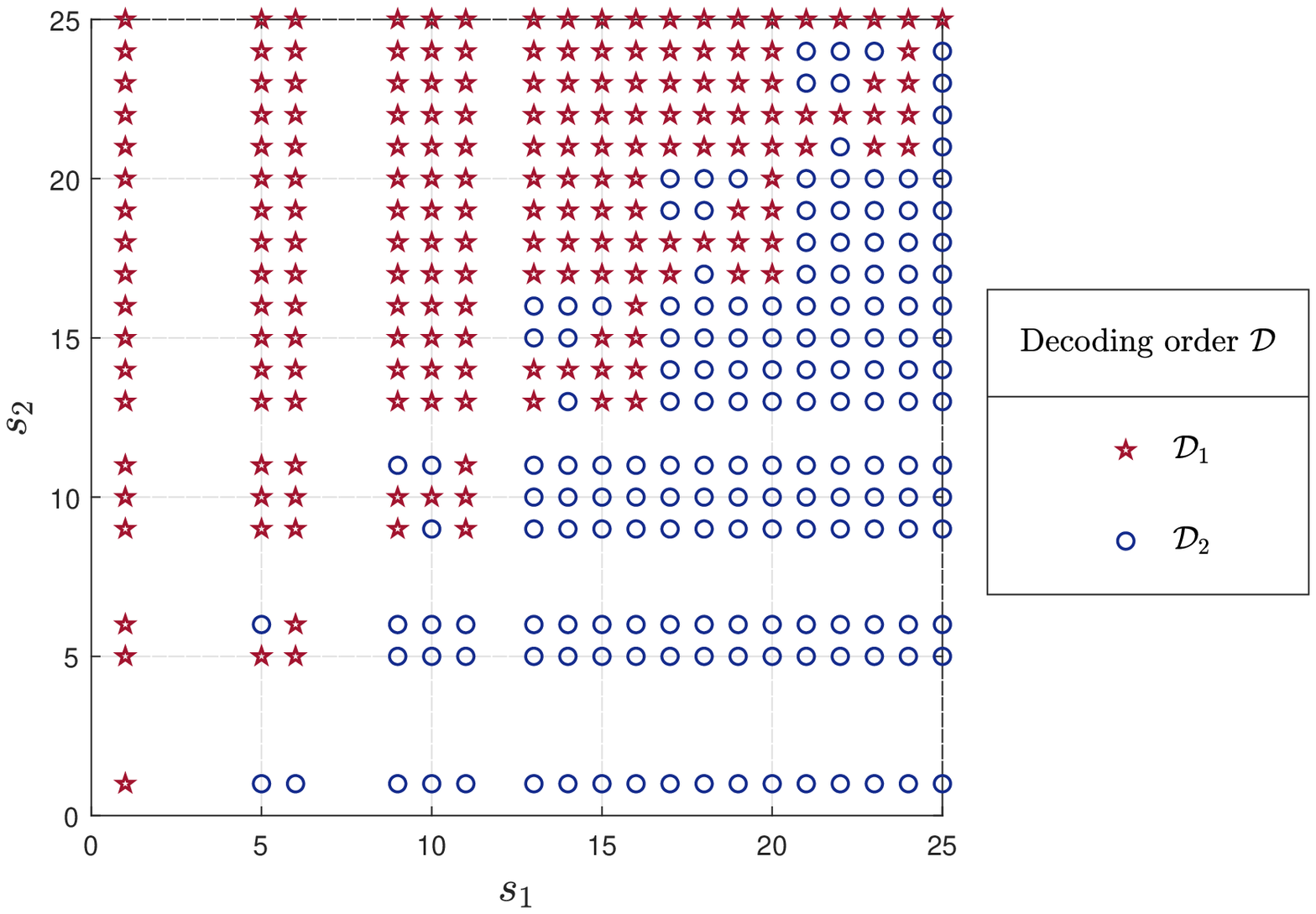}
    \label{fig:OMANOMA5}
    }
    \subfigure[Associated power control policies.]{
    \includegraphics[width=0.47\textwidth]{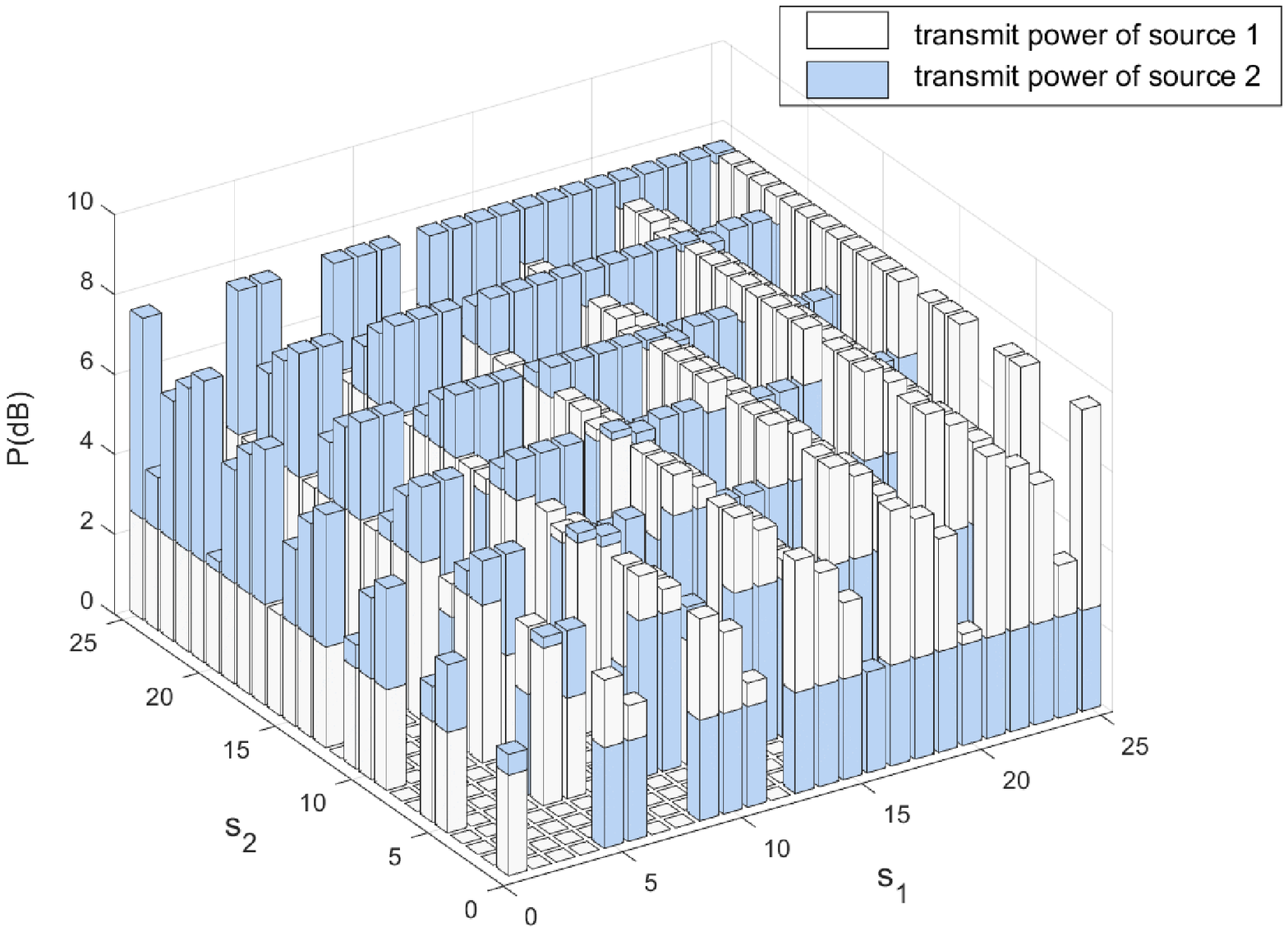}
    \label{fig:Powerallocation}
    }
    \caption{Performance in NOMA when $\overline{P}$ = 5 dB.}
    \label{fig:Performance in NOMA}
\end{figure}

In Fig. \ref{fig:Performance in NOMA}, we plot the optimal decoding order and associated power control policies obtained from Algorithm 2 with $\overline{P}$ = 5 dB, respectively. The state space $\mathcal{S}$ is denoted as $(s_1;s_2)$ where the value of state $s_n$ is mapped to $M\times(\Delta_n- 1)+m_n,\forall n$. Here we truncate a portion of the complete state space to facilitate the analysis and some illogical points have been removed. In Fig. \ref{fig:OMANOMA5}, the distribution of decoding order is symmetrical diagonally, except that there are some reasonable fluctuations in the diagonal. Here we can see that when $s_1$ is large and $s_2$ is small, i.e. in the lower right area of the figure, optimal policy prefers to choose decoding order $\mathcal{D}_2$, which means that the receiver first decodes the message from source 2 such that the message from source 1 sees no interference and attain successful transmissions with higher probability to lower the AoI of source 1, and hence the weighted sum AoI is accordingly reduced. In Fig. \ref{fig:Powerallocation}, we plot the associated power control policies for the two sources. It is interesting that whenever one source is decoded last, the transmit power levels are higher, akin to the opportunistic transmission policies.

\begin{figure}
    \centering
    \includegraphics[width=0.5\textwidth]{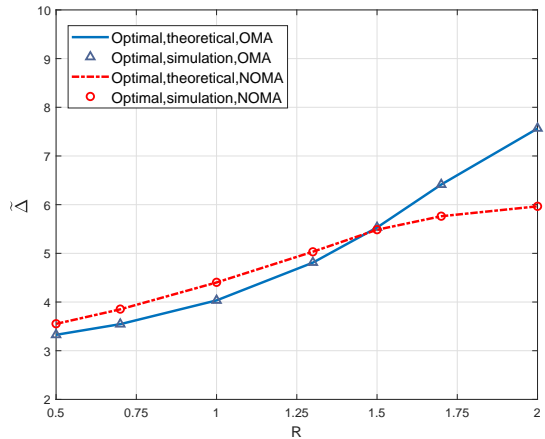}
    \caption{ AoI performance versus packet size $R$ for OMA and NOMA , $\overline{P}$ = 5 dB.}
    \label{fig:OMANOMA6}
\end{figure}

In Fig. \ref{fig:OMANOMA6}, we plot AoI performance versus packet size $R$ for OMA and NOMA. Compared to OMA, NOMA has a better performance when packet size $R$ is larger. In other words, NOMA is better suited for transmitting large update packets, while OMA is reasonable for sending small data packets. It is mainly attributed to the features of NOMA, which allows a large number of accesses and enables resource reuse at the expense of high device complexity. Besides, the large difference in transmit power between sources is key to the fundamental property of the SIC technique, since a larger value of $R$ implies a greater effect on the power values of the action set, and therefore a larger range of power variations available to the sources, which can evidenced in the power expressions (\ref{NOMA Action1 X}) and (\ref{NOMA Action2 X}).
\subsection{Performance in an Unknown Environment}

For the sake of performance comparison with the known environment, we set $R = 1.5$ in the following according to the Fig. \ref{fig:OMANOMA6} in which scenario they have similar age performance.

\begin{figure}
    \centering
    \includegraphics[width=0.5\textwidth]{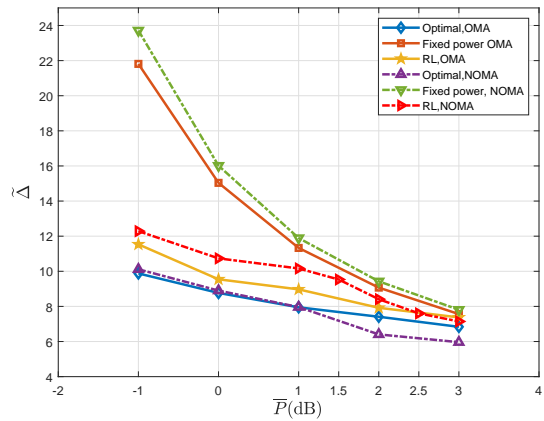}
    \caption{ AoI performance under different schemes.}
    \label{fig:RL}
\end{figure}

In Fig. \ref{fig:RL}, we plot the weighted sum average AoI under different schemes as the average transmission power increases. We notice that the weighted sum AoI derived from the RL algorithms achieve near-optimal performance for no matter OMA or NOMA with no significant deterioration in performance compared to the VIA (i.e., "optimal" in the figure), indicating the validity of the reinforcement learning algorithm. On the other hand, the RL algorithms more markedly reduce the age of information compared to the the policies with fixed power, even if no a priori information is provided, confirming the advantage of the power control policy.

\begin{figure}
    \centering
    \subfigure[Learning process in OMA.]{
    \includegraphics[width=0.47\textwidth]{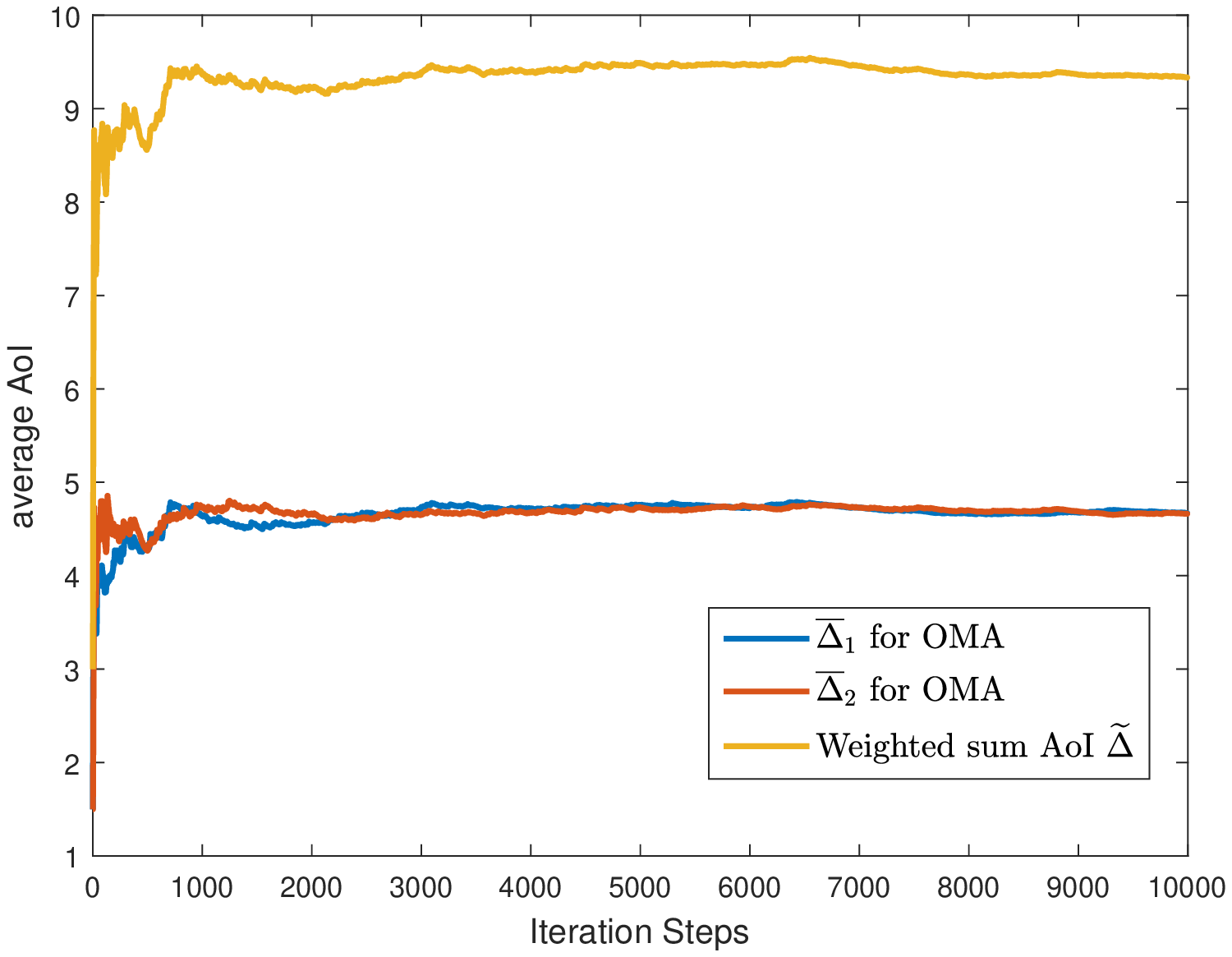}
    \label{fig:OMA_iteration}
    }
    \subfigure[Learning process in NOMA.]{
    \includegraphics[width=0.47\textwidth]{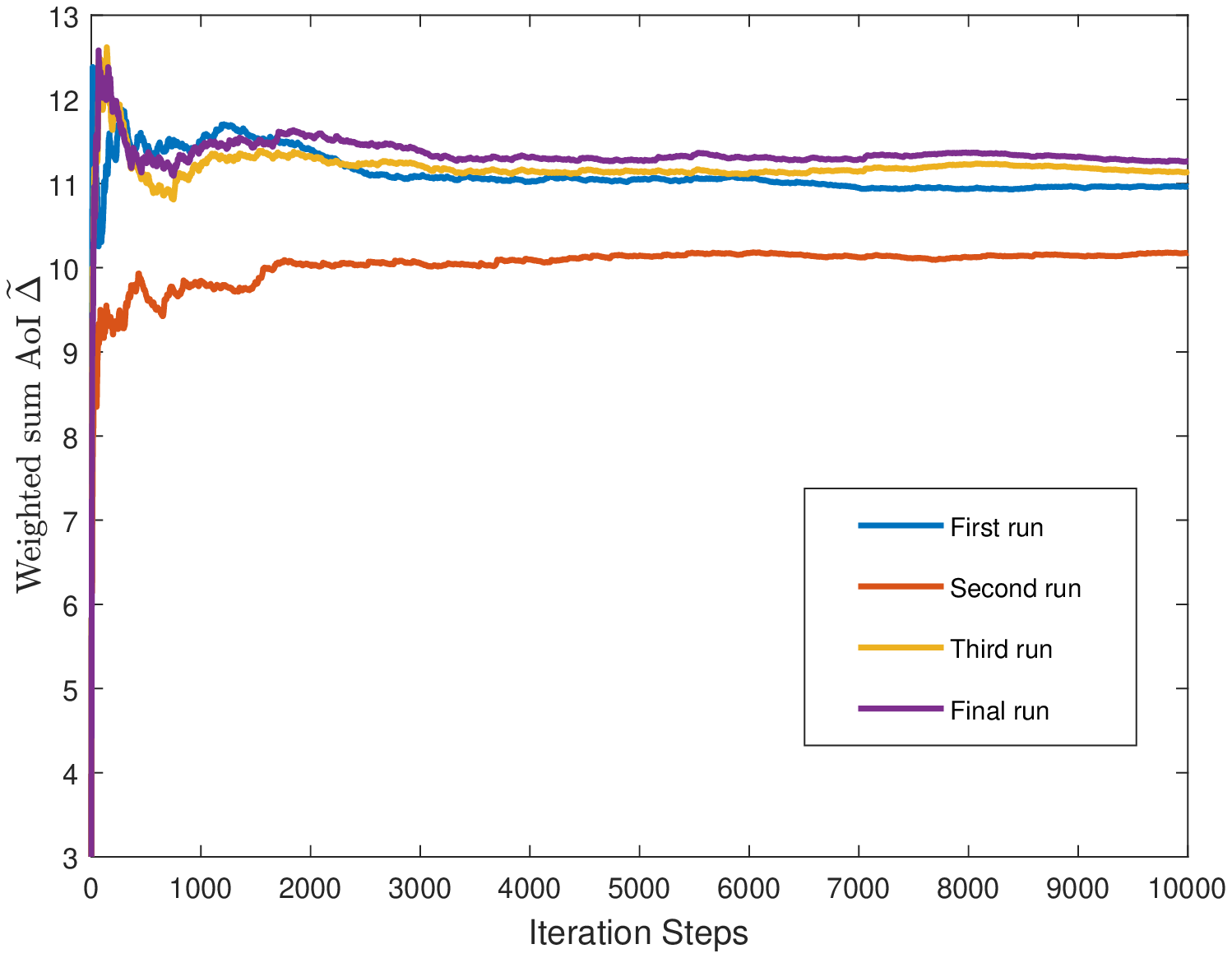}
    \label{fig:NOMA_iteration}
    }
    \centering
    \caption{Evolution of the average age during the learning process when $\overline{P}$ = 0 dB.}
    \label{fig:Rl Performance in NOMA}
\end{figure}

In Fig. \ref{fig:Rl Performance in NOMA}, we investigate the evolution of the average age for the OMA and NOMA cases during the learning process in one episode. As we can see from this figure that in the early period of the learning process, we can intuitively notice a considerable ebb and flow in the value of the average age, which indicates that the Q-learning with $\epsilon-$greedy exploration algorithm we adopted encourages exploration to seek new and potentially better actions in early learning. And we can see that when we have enough exploration, the policy asymptotically becomes conservative and greedy, and the curves are also getting flattened. As shown in Fig. \ref{fig:OMA_iteration}, average AoI $\overline{\Delta}_1$ and $\overline{\Delta}_2$ slowly perform almost identically, except for some differences at the beginning, when they are in the exploration period. While in the NOMA case, some undesired outcomes will occasionally occur depending on the channel realizations just like the second run in the Fig. \ref{fig:NOMA_iteration}. Consequently, we need to make several runs of the simulation to guarantee the reliability of the algorithm by making sure that the difference between the weighted sum average AoI $\widetilde{\Delta}$ obtained from the two consecutive training processes is less than a given threshold as described in Algorithm 4.

\begin{figure}
    \centering
    \subfigure[Initial.]{
    \includegraphics[width=0.47\textwidth]{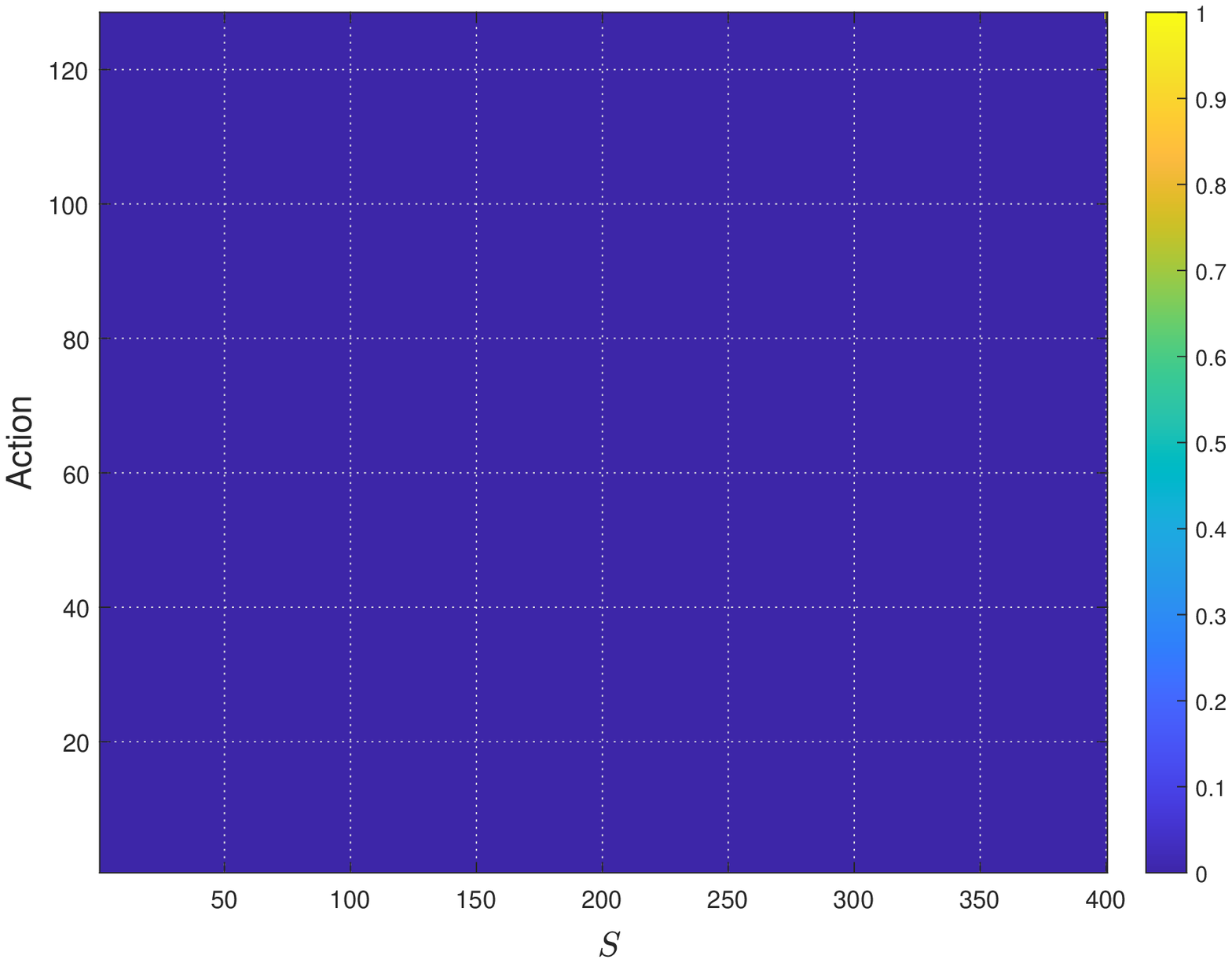}
    \label{fig:initial}
    }
    \subfigure[During learning. $\overline{P}$ = -1 dB.]{
    \includegraphics[width=0.47\textwidth]{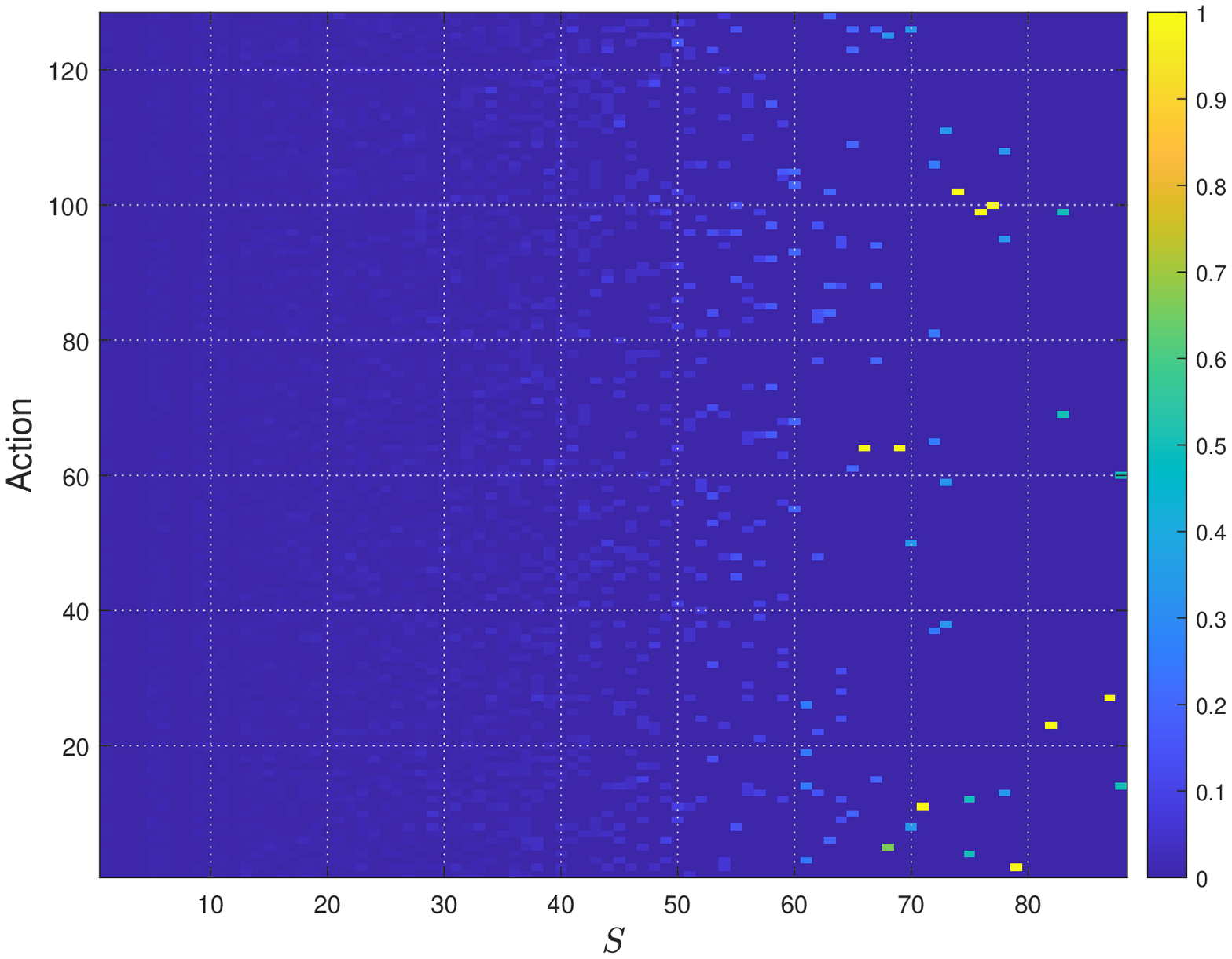}
    \label{fig:-1_s1_imgasc_initial}
    }

    \subfigure[After learning. $\overline{P}$ = -1 dB]{
    \includegraphics[width=0.47\textwidth]{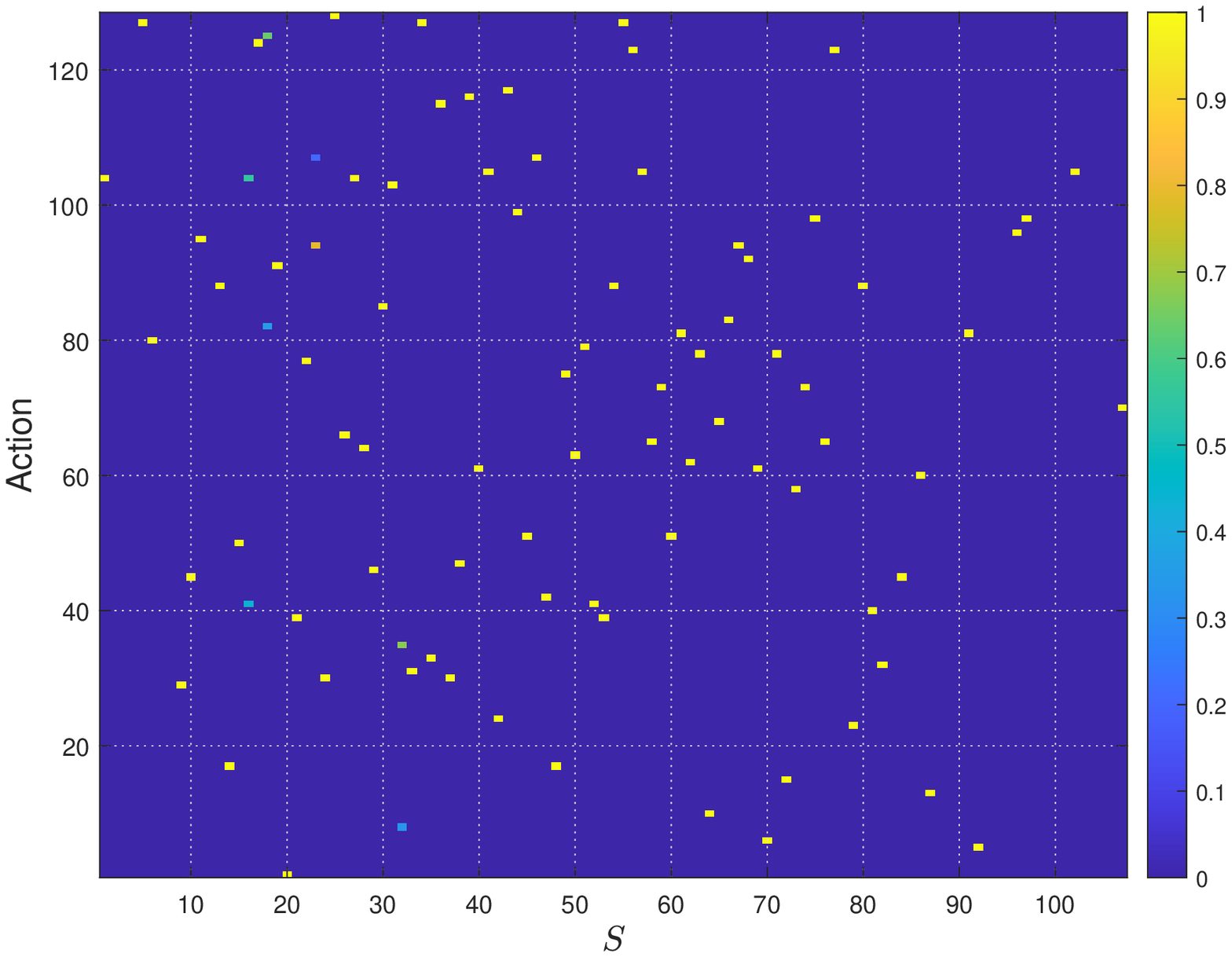}
    \label{fig:-1_s1_imgasc}
    }
    \subfigure[After learning. $\overline{P}$ = 3 dB]{
    \includegraphics[width=0.47\textwidth]{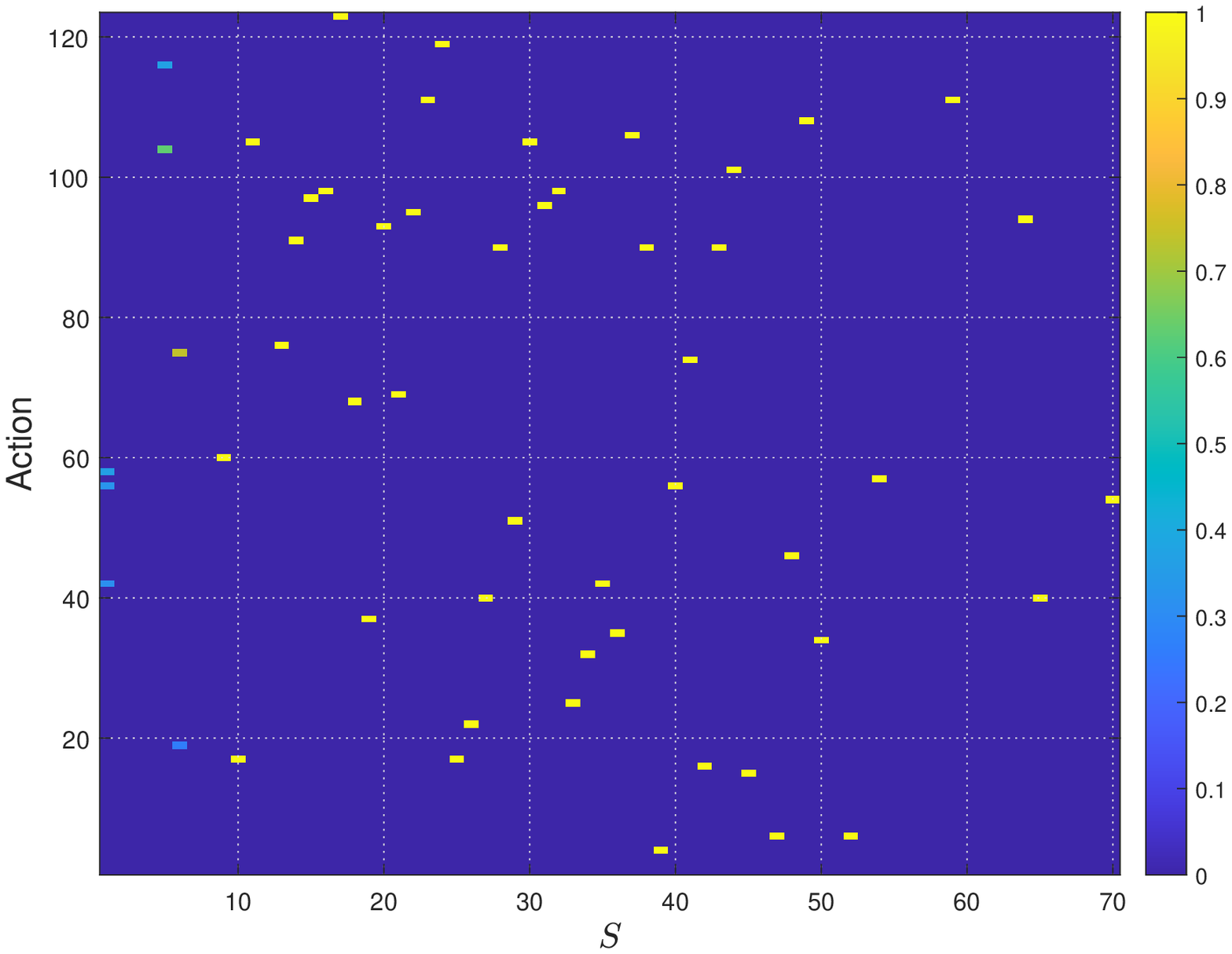}
    \label{fig:3_s1_imgasc}
    }

    \centering
    \caption{Demonstration of action probabilistic selection on the state-action space during the learning process. The results are derived after $10^4$ time steps.}
    \label{fig:action probabilistic selection}
\end{figure}

In Fig. \ref{fig:action probabilistic selection}, we plot the action probabilistic selection under different average power constraints using NOMA scheme. We set the selection probability of all actions at the beginning to zero as in Fig. \ref{fig:initial}. And we can see that except that agents prefer to try some new and potentially good actions at early states, agents will always select the unique and optimal action after enough learning and exploration, confirming the convergence of our algorithms. By comparing Fig. \ref{fig:-1_s1_imgasc} and \ref{fig:3_s1_imgasc}, when the sources are equipped with less power, i.e., -1 dB, the policy is going to visit more states because of higher error transmission probability due to the restricted power set.

\begin{figure}
    \centering
    \includegraphics[width=0.47\textwidth]{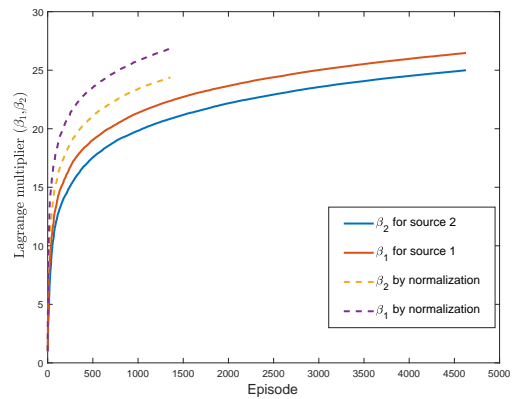}
    \caption{ The time consumption to obtain the Lagrangian multipliers. Each episode consists of $10^4$ time steps.}
    \label{fig:beta}
\end{figure}

In Fig. \ref{fig:beta}, we show the convergence paths to find the optimal Lagrange multipliers $\beta_1$ and $\beta_2$, and each episode consists of $10^4$ time steps. It is apparent from this figure that through normalization, we speed up the convergence of the algorithm and save time greatly without changing the performance significantly in the process of seeking Lagrange multipliers.

\section{Conclusions}
In this paper, we have analyzed the AoI performance optimization on different multiple access schemes (i.e., OMA and NOMA) employing optimal power allocation policy, where different transmit power level is selected according to different channel state. We have assumed that multiple independent power-constraint sources send update packets to a common receiver and the maximum number of retransmission rounds can not exceed $M$ times. Thus, considering the two transmission schemes, we have formulated CMDP problems to minimize the weighted sum average AoI under average power constraints, respectively. We have resorted to the Lagrangian method to transfer CMDP problems to equivalent MDP problems and obtained the corresponding Value Iteration Algorithms to derive the power allocation policy. However when the environment information is not known as a priori, these analyses are no longer applicable and we have proposed online reinforcement learning algorithms. Through numerical results, we have verified that the proposed optimal policies reduce the weighted sum average AoI more significantly compared to fixed power policy and demonstrated that NOMA is more suitable for transmitting large update packets due to higher spectral efficiency introduced. What's more, Q-learning with $\epsilon$-greedy exploration algorithms achieve near-optimal age performance and save a lot of time by normalization.

\end{document}